\documentclass[a4paper]{article}
\pdfoutput=1
\usepackage{pdfpages}

\usepackage{icrc2013}
\usepackage{amsmath}
\usepackage[english]{babel}
\usepackage[utf8]{inputenc}
\usepackage[T1]{fontenc}
\usepackage{cite}
\usepackage{url}
\usepackage{hyperref}
\hypersetup{
  colorlinks=true,
  linkcolor=blue,
  citecolor=blue,
  urlcolor=blue,
}

\newcommand{\nosection}[1]{%
  \refstepcounter{section}%
  \addcontentsline{toc}{section}{\protect\numberline{\thesection}#1}%
  \markright{#1}}

\title{Dark Matter, Cosmology, and Fundamental Physics with HAWC:\\Contributions to ICRC 2013}

\authors{
{\bf The HAWC Collaboration:}\\
A.~U.~Abeysekara$^{a}$,
R.~Alfaro$^{b}$,
C.~Alvarez$^{c}$,
J.~D.~{\'A}lvarez$^{d}$,
R.~Arceo$^{c}$,
J.~C.~Arteaga-Vel{\'a}zquez$^{d}$,
H.~A.~Ayala Solares$^{e}$,
A.~S.~Barber$^{f}$,
B.~M.~Baughman$^{g}$,
N.~Bautista-Elivar$^{h}$
% ...
}

\afiliations{
{\center
$^{a}$ Department of Physics and Astronomy, Michigan State University, East Lansing, MI, USA\\
$^{b}$ Instituto de F{\'\i}sica, Universidad Nacional Aut{\'o}noma de M{\'e}xico, Mexico D.F., Mexico\\
$^{c}$ CEFyMAP, Universidad Aut{\'o}noma de Chiapas, Tuxtla Guti{\'e}rrez, Chiapas, Mexico\\
$^{d}$ Universidad Michoacana de San Nicol{\'a}s de Hidalgo, Morelia, Mexico\\
$^{e}$ Department of Physics, Michigan Technological University, Houghton, MI, USA\\
$^{f}$ Department of Physics and Astronomy, University of Utah, Salt Lake City, UT, USA\\
$^{g}$ Department of Physics, University of Maryland, College Park, MD, USA\\
$^{h}$ Universidad Politecnica de Pachuca, Pachuca, Hgo, Mexico\\
% ...
}
}

\abstract{
  The High-Altitude Water Cherenkov Gamma Ray Observatory (HAWC) is designed to
  perform a synoptic survey of the TeV sky.  The high energy coverage of the
  experiment will enable studies of fundamental physics beyond the Standard
  Model, and the large field of view of the detector will enable detailed
  studies of cosmologically significant backgrounds and magnetic fields.  We
  describe the sensitivity of the full HAWC array to these phenomena in five
  contributions shown at the 33$^\text{rd}$ International Cosmic Ray Conference
  in Rio de Janeiro, Brazil (July 2013).
}

\keywords{dark matter, WIMP, Q-ball, Lorentz invariance violation, primoridial
black holes, intergalactic magnetic fields}

\begin{document}
\maketitle

\pagestyle{plain}
\pagenumbering{arabic}

\clearpage

% Contents
\newpage
\onecolumn{
  \tableofcontents
}

% Q-balls
\newpage
\setcounter{section}{0}
\nosection{Searching for Q-balls with the High Altitude Water Cherenkov
Observatory\\
{\footnotesize\sc Peter Karn, Patrick Younk}}
\setcounter{section}{0}
\setcounter{figure}{0}
\setcounter{table}{0}
\setcounter{equation}{0}

\title{Searching for Q-balls with the High Altitude Water Cherenkov Observatory}

\shorttitle{Searching for Q-balls with HAWC}

\authors{ Peter Karn$^{1}$, Patrick Younk$^{2}$ for the HAWC Collaboration$^{3}$ }

\afiliations{
  $^1$ Dept. of Physics and Astronomy, University of California, Irvine \\
  $^2$ Los Alamos National Laboratory \\
  $^3$ For a complete author list, see the special section of these proceedings 
}

\email{pkarn@uci.edu}

\abstract{The High Altitude Water Cherenkov (HAWC) Observatory is a gamma-ray
experiment currently under construction at Sierra Negra in Mexico. When complete
it will consist of a 22,000 square meter array of 300 water Cherenkov detectors.
Although HAWC is designed to study gamma rays from galactic and extra-galactic
sources, the large volume of instrumented water (each tank holds $\sim$200,000
liters) gives the opportunity to search for more exotic species. One such
target, predicted by several varieties of supersymmetric theory, is the Q-ball.
Q-balls are very massive, subrelativistic particles that can have a large baryon
number and can be stable since their creation in the early universe. They are
also an appealing candidate for the dark matter of the universe, but their large
masses must mean their flux is very low. HAWC has a flexible data acquisition
system which, with a dedicated trigger algorithm for non-relativistic species,
allows a search for Q-balls traversing the detector. The trigger algorithm and
sensitivity are presented here.
}

\keywords{Q-ball, sensitivity, HAWC}

%\begin{document}
\maketitle

\section*{Introduction to Q-balls}
\label{intro}

The Q-ball is an exotic form of matter predicted by supersymmetric theory
\cite{coleman}.  Named for its ability to carry a large baryon number and its
spherical shape, a Q-ball is a condensate of the scalars available
in supersymmetry, i.e., squarks, sleptons, and Higgs fields.  The mass of a
Q-ball is dependent on its baryon number \cite{kkst}:
\begin{equation}
  M_{Q}=\frac{4\pi\sqrt{2}}{3}M_{S}Q^{3/4}
  \label{masseq}
\end{equation} 
where $M_{S}$ is the energy scale at which supersymmetry is broken (an unknown
parameter but generally assumed to be within an order of magnitude of 1 TeV) and
Q is the baryon number carried by the Q-ball.  The interaction cross section is
approximately equal to the physical size of the Q-ball:
\begin{equation}
  \sigma=\frac{\pi}{2}M_{S}^{-2}Q^{1/2}.
  \label{crosssectioneq}
\end{equation}
One estimate \cite{laine} based on the theory of their creation (discussed
below) and taking into account some cosmological requirements finds the most
likely baryon number for Q-balls around today to be around $10^{24}$ plus or
minus a few orders of magnitude.  Given Eqs. \ref{masseq} and
\ref{crosssectioneq} with $M_{S}$ = 1 TeV, these Q-balls would have a mass of
about $10^{19}$ TeV (or 20 micrograms) packed into the size of an atom.  

The Q-ball is an example of what is known in field theory as a non-topological
soliton.  This means that it is a stable field configuration that has some
conserved charge, in this case baryon number.  It could decay into normal
baryons but will not if it is energetically disfavorable to do so.  Considering
Eq. \ref{masseq}, it is clear that as baryon number increases, at some point the
Q-ball will be less massive than the same baryon number stored in protons.  A
simple calculation gives an estimate of the stability condition:
\begin{equation}
  Q>1.6\times10^{15}\left(\frac{M_{S}}{1~\textrm{TeV}}\right)^{4}.
  \label{stabilitycondition}
\end{equation}
Therefore, if $M_{S}$ = 1 TeV, for example, a Q-ball must have a mass over
$10^{15}$ GeV (or a 25 mb cross section) to be stable.  If Q-balls meet this
condition then they would be observable in the present day and contribute to the
dark matter.  Their large masses mean they could exist in the galactic halos we
observe for dark matter because they would not have had time to relax into the
disk.  Just like models of halos made of WIMPs, the Q-balls bound to the galaxy
should have a velocity distribution at the radius of Earth which can be modeled
by a maxwellian centered at 230 km/s \cite{dmvelocity}.  However, to match the
observed density of dark matter, their flux must be very small, so only a
detector with a very large area would have any chance of detecting them.

The theory of the creation of Q-balls in the early universe is called
Affleck-Dine baryogenesis \cite{affleck,dine}.  The Affleck-Dine mechanism is a
natural consequence of supersymmetry and inflation.  This theory states that
after inflation, the universe consisted of a single, nearly uniform scalar
condensate: one enormous Q-ball.  Over time, small inhomogeneities caused the
Affleck-Dine condensate to become unstable and fragment.  It is thought that the
Q-balls thus produced should be of a fairly narrow band of masses
\cite{dmqballs}.  The fate of such Q-balls depends on the model of supersymmetry
and cosmology considered.  They could then decay into ordinary baryons and
neutralinos or some could decay while others remained stable.  Either way, by
providing a common origin for baryons and dark matter, Affleck-Dine baryogenesis
gives an explanation for the similarity between the abundances of these two
classes of matter in our universe.  

Q-balls can answer two of the most important questions in astrophysics: ``What
is the dark matter?'' and ``What is the origin of the baryon asymmetry?''
Detection of Q-balls is the only known way to confirm that Affleck-Dine
baryogenesis is correct, and obviously it is also the only way to identify them
as dark matter.  Therefore, searches for Q-balls with the most suitable
detectors available today are warranted.

\section*{Q-ball Interactions}

Q-balls are required to be so massive that they would not lose an appreciable
amount of their kinetic energy even if they passed directly through a star.
Neutron star material is dense enough to stop a Q-ball, and some restrictions on
Q-ball theory based on the existence of undisturbed neutron stars have been
calculated \cite{astrobounds}.  However, a Q-ball passing through ordinary
matter would interact frequently and strongly.  The interactions are a direct
result of the fact that the vacuum inside the surface of a Q-ball is distinct
from the rest of the universe, i.e., the rules of particle physics are
fundamentally different.  The symmetry that governs the strong force, known as
$SU(3)_{color}$, is broken in the interior of a Q-ball.  Therefore, when a
Q-ball is incident upon a nucleon, the nucleon and its constituents no longer
feel the strong force and the nucleon is dissociated into quarks.  The
$U(1)_{baryon}$ symmetry is also broken inside a Q-ball.  As a result, the
quarks just released are temporarily indistinguishable from antiquarks.  They
then have a probability of order one to be reflected from the Q-ball surface as
antiquarks with the Q-ball absorbing the difference in baryon number.  If the
interaction was between a Q-ball and a single, lone nucleon, the quarks would
then hadronize into pions and carry away the binding energy of the nucleon,
about 1 GeV \cite{kusenko05}.

In bulk matter, the additional particles available will make the interaction
more complicated.  If the bulk matter is water, for example, the important
``scattering center'' to consider is an oxygen nucleus.  In an interaction with
such a nucleus the Q-ball would destroy nucleons as described above and create
antiquarks.  The antimatter would then annihilate with the other nucleons in the
nucleus.  The end result is equivalent to the annihilation products of the
sixteen nucleons in an oxygen nucleus.  To first approximation this would be on
average 40 pions each carrying around 400 MeV, though of course other particles
can and will be produced.  

This large deposition of energy is more than enough to be detected by modern
experiments if it occurs inside our detectors and technical issues, such as
background suppression, can be overcome.  As mentioned above, relic dark matter
Q-balls would be traveling at subrelativistic speeds.  Then the signature of a
Q-ball is a slow moving particle depositing a lot of light as it traverses a
detector.  A small Q-ball with a cross section of 25 mb (the smallest stable
Q-ball) would deposit 1.34 GeV/m while a larger one with a cross section of 600
barns (more likely) would deposit around 32 TeV/m.

\section*{Previous Measurements} 

A number of experiments have searched for evidence of Q-balls and established
upper limits on their flux on Earth.  The current state of the field is defined
by results from two experiments: Super-Kamiokande II \cite{superk} and MACRO
\cite{macro}. 

The Super-Kamiokande detector is broken into an inner detector containing the
majority of its fiducial mass and instrumentation and an outer detector used to
veto cosmic ray muons.  The measurement by Super-Kamiokande, which used 541.7
days of live time, was made by requiring at least two Q-ball interactions in the
inner detector (with appropriate cuts on timing and energy deposited) while also
requiring that few PMTs were hit in the outer detector to remove muon events.
This approach allowed them to set a limit in the range of cross sections from
0.02 to 200 mb.  The muon cut limits their ability to probe higher cross
sections because a Q-ball of high cross section would deposit too much energy in
the outer detector.  In addition, the way Q-ball interactions were modeled for
the Super-Kamiokande analysis differs from that described above.  They
considered only the interaction between a Q-ball and a single nucleon (as
opposed to an oxygen nucleus), which would produce between one and three pions
carrying a total of 1 GeV.  They then modeled the water as having a uniform
density of nucleons.  Therefore, the energy deposited per unit length would be
the same as the model for HAWC, but the energy released per interaction and the
frequency of interactions are different by a factor of 16.

The limit due to MACRO data is actually a reinterpretation of their limit on
magnetic monopoles.  This was done in the Super-Kamiokande paper to have a point
of comparison for their limit, and they did the same with an earlier magnetic
monopole limit from Kamiokande.  In their model, a Q-ball interaction would have
a very similar signature to a magnetic monopole catalyzing nucleon decay, and
this measurement and analysis differs from HAWC's for similar reasons.
Nonetheless, the above two limits are presented in Fig. \ref{sensitivity} below
for comparison to HAWC's sensitivity.  Regardless of differences in interaction
  models, it is reasonable to expect that HAWC could set a stronger limit than
  Super-Kamiokande or MACRO because of its greater size (60 ktons of water
  compared to Super-Kamiokande's 50 ktons) as well as it's flat detector
  geometry, which gives more effective area for a given volume.

It should be mentioned that a third limit on Q-ball flux was made, but never
published, by the AMANDA experiment.  The results are recorded in a Ph. D.
thesis \cite{amanda}.  This measurement, too, differs from HAWC's in that
only Q-ball velocities greater than 900 km/s were considered, which is too fast
for relic dark matter bound to our galaxy.  Finally, it is probable that
  IceCube, as the successor to AMANDA, could perform a search for Q-balls but
  they have not done so as of this writing. 

\section*{The High Altitude Water Cherenkov Observatory}

The High Altitude Water Cherenkov (HAWC) observatory is a gamma-ray experiment
currently being constructed in the mountains near Puebla, Mexico.  As the
successor the the Milagro experiment, it is a second generation extensive air
shower array.  It is being built at an altitude of 4,100 meters a.s.l. and when
completed will span an area of $\sim$22,000 $m^2$.  HAWC will be most sensitive 
to gamma rays from 100 GeV up to 100 TeV.  With its wide field of view, HAWC will 
scan about half the sky each day.  In addition, the water cherenkov technique 
allows HAWC to have a high duty cycle, \textgreater90\%.  Therefore, HAWC is 
especially well suited for galactic surveys and galactic and extra-galactic 
transient detection, but will probe acceleration mechanisms of all gamma ray 
sources in its energy range.

The completed array will consist of 300 optically isolated water cherenkov
detectors (WCDs), each holding $\sim$200,000 liters of water.  The WCDs are
composed of cylindrical steel tanks measuring 7.3 meters in diameter and filled
to a height of 4.5 meters.  A WCD is instrumented with four photomultiplier
tubes on the bottom looking up: one Hamamatsu R7081 high quantum efficiency PMT
in the center and 3 Hamamatsu R5912 PMTs positioned halfway to the edge of the
WCD.  HAWC's modular design allows the experiment to be operated well before
construction is complete.  HAWC is taking data now with 95 WCDs and is scheduled
to be completed in the summer of 2014.

The data acquisition system consists of custom front-end electronics boards
which are being reused from the Milagro experiment.  The front end boards are
used to shape the PMT pulse and digitize the crossing times of the pulse over
two fixed voltage thresholds.  The duration for which each pulse was over the
two thresholds, time over threshold (TOT), is used as a proxy for pulse height,
as the entire waveform is not digitized.  The edge times are fed into CAEN
V1190 TDCs and processed by the on-site computing farm.  
                                                                            
One of the most critical aspects of HAWC's design that allows for the search for
Q-balls and other exotic species, is the software trigger.  HAWC's 1200 PMTs
will generate an estimated 500 MB/s of raw data, which must be reduced before it
can leave the site.  Only $\sim$20 MB/s will be permanently recorded.  To accomplish
this, HAWC employs a highly flexible and totally software-based trigger.
Although the standard trigger for air showers is based on multiplicity of PMTs
hit in a short time window (on the order of a microsecond), other triggers can
be run in parallel.  For example, fast reconstruction algorithms can be run over
signals with smaller multiplicity to recover smaller air showers and improve low 
energy sensitivity.  Custom triggers for exotic species can also be utilized.  
Since Q-balls are subrelativistic, the time scale of their interaction must be 
very different from anything else that HAWC was designed to detect.  Therefore, 
a dedicated Q-ball trigger is necessary.
 
\section*{Trigger Algorithm}

HAWC's trigger algorithm for Q-balls works as follows:  We require that at least
three out of the four PMTs in a WCD are hit above a threshold of 10
photoelectrons within 50 ns.  This is called a tank hit and represents the
potential detection of one Q-ball interaction in the water.  However, that
signature alone is not sufficient because it could be mimicked by a muon
striking a WCD and depositing a large amount of energy.  So, we must eliminate
that background by requiring multiple interactions.  Many orders of magnitude
are allowed for the size of a Q-ball, but with a high enough cross section
($\sim$100 mb), a Q-ball passing through a WCD would interact multiple times.
Several muons can still hit with timing that mimics a Q-ball signal, but of
course it would be rarer.  Since the times of the hits are recorded and the size
of the WCD is known, we have an estimate of the maximum speed of the potential
particle.  We can then cut on this value to match the expected velocity
distribution for Q-balls.  After adjusting for the motion of the earth, the
average measured velocity should be about 320 km/s.  Then the time for a Q-ball
to cross a 7 meter WCD is on average is 22 us.  We want to account for Q-balls
with low velocities, so we look in a sliding time window of 86 us, which should
save over 98\% of the expected distribution.  

The $n$-fold coincidence rate of a single detector with a hit rate of $r$ in a
time interval $\Delta t$ is given by Poisson statistics: 
\begin{equation} 
  R_{n} = \frac{r\cdot (r\Delta t)^{n-1}\cdot e^{-r\Delta t}}{(n-1)!} 
  \label{triggerrate} 
\end{equation} 
The rate of muons incident upon a WCD is on the order of 10 kHz.  It is
difficult to accurately estimate what fraction of those muons would meet the
conditions for a tank hit, but the measured rate of such tank hits is about
1,850 kHz per tank.  Then using Eq. \ref{triggerrate} gives an estimate of the
background trigger rate as a function of the number of hits required.  To reach
an acceptable trigger rate we require at least six hits, giving a calculated
trigger rate of 0.0013 Hz per tank.  This matches well with the measured rate of
0.0014 Hz.

All PMT hits in the entire array are saved for several milliseconds around a
Q-ball trigger so that further reconstruction can be done thoroughly.  For
example, if some of the hits causing a Q-ball trigger were demonstrably
associated with an air shower, they could be discarded.  Further, the timing and
reconstructed locations of the interactions can be checked to see if they are
consistent with the slow, straight track that a Q-ball should have.  

\section*{Saturation Trigger}

There is a limitation in the effectiveness of the above trigger algorithm for
Q-balls of very high cross section due to the fact that such a Q-ball passing
through a WCD would be creating so much light that it would saturate the
electronics.  Specifically, the interactions, and therefore PMT pulses, would be
happening so fast they would begin to have a good chance of overlapping above
$\sim$2 barns.  As mentioned above and shown below, cross sections in the range
of hundreds to thousands of barns seem to be the most theoretically motivated
sizes to look for.  Therefore, although it is clear that the previously
described trigger will not work here, we would like to be able to set a limit at
high cross sections if possible.  To do so, we are currently evaluating the
possibility of triggering on the signature of four saturated PMTs over a period
of tens of microseconds.  We can mimic the fast pulses with a high frequency
laser and study the waveforms produced as well as the digital signal from the
front-end electronics.  This scheme should work unless the saturation mode of
the PMTs is indistinguishable from normal operation.  Again, due to the software
trigger, all that would be needed is another piece of code to search for
appropriate signature.

\section*{Sensitivity}

\begin{figure}
  \centering                                              
  \includegraphics[width=0.5\textwidth]{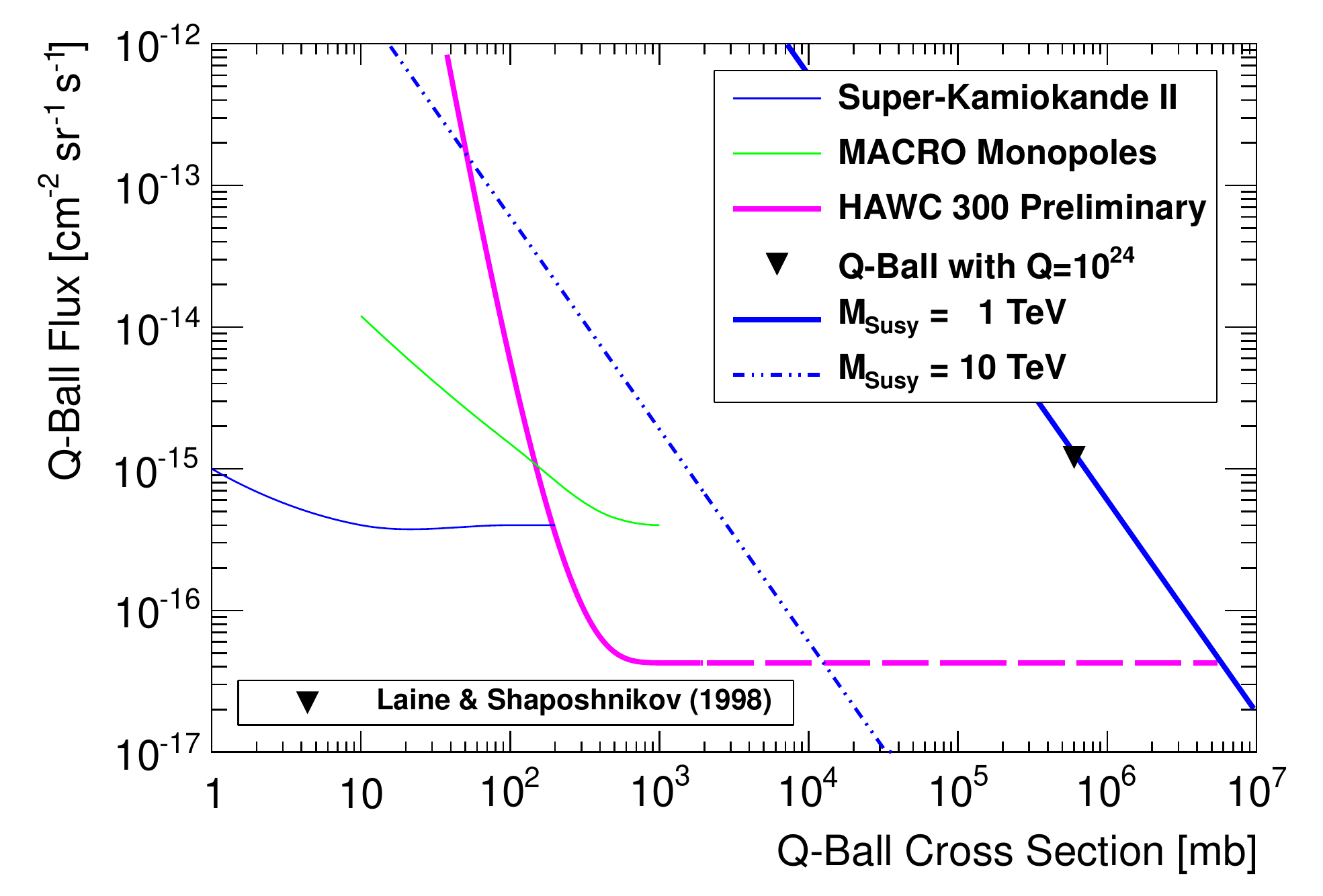}
  \caption{Estimated flux upper limit for HAWC with one year of the full array
  taking data assuming no background, as compared to current limits.  The solid
  line indicates what can definitely be done with the current Q-ball trigger
  algorithm while the dashed line indicates cross sections which could be
  accessible if the saturation trigger is viable.  The blue lines on the right
  show expectations from theory: if one assumes all the dark matter is composed
  of Q-balls, then the estimated dark matter density, velocity distribution, and
  Q-ball mass can be used to calculate the expected flux.  The Q-ball mass can
  be written as a function of cross section and the scale at which SUSY is
  broken.  Since $M_{SUSY}$ is unknown, the flux vs. cross section is plotted
  for two values for illustrative purposes.  The black triangle represents the
    estimate mentioned in Sec. \ref{intro} where the details of Affleck-Dine
    baryogenesis were considered it was required that the correct baryon
    asymmetry we observe today is generated \cite{laine}.} 
  \label{sensitivity}
\end{figure}

\begin{figure}
  \centering
  \includegraphics[width=0.5\textwidth]{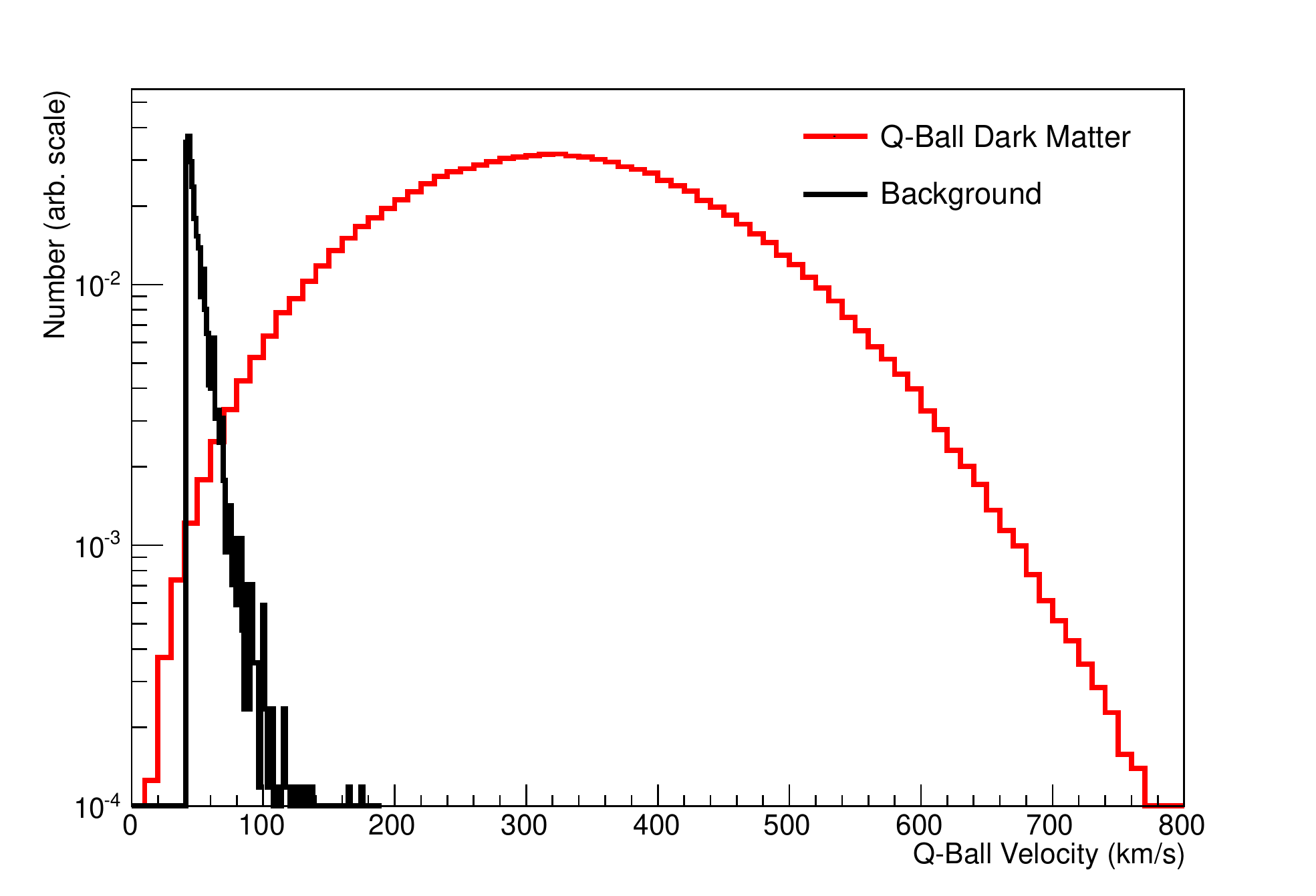}
  \caption{Velocity distributions measured from triggers and expected velocity
  distribution for dark matter Q-balls.  The histograms have been scaled for
  ease of comparison.}
  \label{velplot}
\end{figure}

The sensitivity of HAWC to Q-balls can be estimated in a straightforward manner.
Since a Q-ball can pass through the entire planet without being significantly
affected, HAWC can accept Q-balls from above, below, or the side.  An estimate
of the effective aperture is then given by multiplying the appropriate solid
angles by the areas for the top, bottom, and sides of the HAWC array.  Adding
these up gives an aperture of $(A\cdot \Omega)_{eff} = 172,000~\textrm{m}^2
\cdot \textrm{sr}$.  Using the standard machinery for computing upper limits
\cite{upperlimit}, the one-year sensitivity of HAWC at 90\% confidence level is:
\begin{equation}
  \frac{2.3}{172,000~\textrm{m}^2 \cdot \textrm{sr} \cdot 1~\textrm{yr}} =
  4.26\times 10^{-17}~\textrm{cm}^{-2} \cdot \textrm{sr}^{-1} \cdot
  \textrm{s}^{-1}.
  \label{sensieq}
\end{equation}  
As mentioned above, the standard trigger requires at least six interactions in a
single WCD which is very unlikely for low cross sections.  This means our
sensitivity ``turns on'' when it becomes likely to have at least six
interactions in a few meters of water.  The probability of a Q-ball of cross
section $\sigma$ interacting $N$ times in a distance $x$ is:
\begin{equation}
  P(N)=\frac{(x\sigma n)^N}{N!}\cdot e^{-x\sigma n}
  \label{psubn}
\end{equation}
where $n$ is the number density of water molecules.  The probability of getting
6 or more interactions is then:
\begin{equation}
  P(\ge6)=1-P(0)-P(1)...-P(5).
  \label{pge6}
\end{equation}
Shown in Fig. \ref{sensitivity} is HAWC's sensitivity to Q-balls for one year of
the full array operating, in comparison to established limits and a prediction
from theory.  This curve assumes that all background triggers can be removed
after reconstruction.  
This is a reasonable assumption since there are many quantities to discriminate
with.  Studies of minimum bias data have shown that even the simplest approach,
tightening the cuts in the trigger algorithm, can eliminate the vast majority of
background triggers.  For example, the distribution of maximum velocities for
recorded triggers (assumed to be background) is quite different from the
velocity distribution expected for Q-ball dark matter (see Fig. \ref{velplot}).

\section*{Conclusion}

The High Altitude Water Cherenkov Observatory is taking data now, and will
continue to do so as the array grows over the next year.  HAWC is poised to
provide us with an unprecedented view of the sky in the energy range from 100
GeV to 100 TeV.  The sensitivity and flexibility of HAWC also allows us to study
a number of fundamental physical and cosmological questions.  The Q-ball, as a
dark matter candidate and a potential explanation of the baryon asymmetry, is a
theoretically well motivated target for a search.  HAWC will be able to set the
strongest limit to date over the range of cross sections to which it is
sensitive, and could potentially extend a limit into a region of parameter space
that has never been probed. 

\section*{Acknowledgments}

We acknowledge the support from: US National Science Foundation (NSF); US
Department of Energy Office of High-Energy Physics; The Laboratory Directed
Research and Development (LDRD) program of Los Alamos National Laboratory;
Consejo Nacional de Ciencia y Tecnolog\'{\i}a (CONACyT), M\'exico; Red de
F\'{\i}sica de Altas Energ\'{\i}as, M\'exico; DGAPA-UNAM, M\'exico; and the
University of Wisconsin Alumni Research Foundation.

\clearpage

%\end{document}

% PBH
\newpage
\setcounter{section}{1}
\nosection{HAWC Sensitivity for the Rate-Density of Evaporating Primordial
Black Holes\\
{\footnotesize\sc Tilan Ukwatta, Jane~H. MacGibbon, Daniel Stump, Gus Sinnis,
James~T. Linneman, Kristen Tollefson, A.~Udara Abeysekara, Dirk Lennarz}}
\setcounter{section}{0}
\setcounter{figure}{0}
\setcounter{table}{0}
\setcounter{equation}{0}
%%
% 33nd International Cosmic Ray Conference - 2013 - Rio de Janeiro, Brazil
% Template adapted from the 2011 ICRC template.

%\documentclass[a4paper]{article}
%
%\usepackage{icrc2013}
%\usepackage{amsmath}

%The paper title
\title{HAWC Sensitivity for the Rate-Density of Evaporating Primordial Black Holes}

%The short title to appear at the header of the pages.
\shorttitle{HAWC PBH Sensitivity}

%All paper authors
\authors{
T. N. Ukwatta$^{1}$,
J. H. MacGibbon$^{2}$,
D. Stump$^{1}$,
G. Sinnis$^{3}$,
J. T. Linnemann$^{1}$,
K. Tollefson$^{1}$,
A. U. Abeysekara$^{1}$, and
D. Lennarz$^{4}$
for the HAWC Collaboration.
}

%All the affiliations.
\afiliations{
$^1$ Department of Physics and Astronomy, Michigan State University, East Lansing, MI 48824, USA. \\
$^2$ Department of Physics, University of North Florida, Jacksonville, FL 32224, USA. \\
$^3$ Los Alamos National Laboratory, NM, USA.\\
$^4$ School of Physics and Center for Relativistic Astrophysics, Georgia Institute of Technology, Georgia, USA.
%\scriptsize{
%$^{4}$ now at: Institute of Physics 4. \\
%$^{5}$ now at: Institute of Physics 5.
%}
}

\def\APJ{\emph{ApJ.}}
\def\apj{\emph{ApJ.}}
\def\aap{\emph{A.\& A.}}
\def\apjs{\emph{ApJS}}
\def\apss{\emph{Ap\&SS}}
\def\apjl{\emph{ApJ. Lett.}}
\def\araa{\emph{ARA\&A}}
\def\mnras{\emph{MNRAS}}
\def\nat{\emph{Nature}}
\def\NAT{\emph{Nature}}
\def\pre{\emph{Phys. Rev. E}}
\def\prd{{\em Phys. Rev.} D}
\def\pasj{\emph{PASJ}}

%email address of the contact person
\email{tilan.ukwatta@gmail.com}

%The abstract.
\abstract{
Primordial black holes (PBHs) are hypothetical black holes that may have formed from extreme densities of matter present during the early universe. Hawking showed that due to quantum-gravitational effects, a black hole possesses a temperature inversely proportional to its mass and emits with a thermal spectrum all species of fundamental particles. PBHs with initial masses of $\sim 5.0 \times 10^{14}$ g should be expiring now with bursts of high-energy particles, including gamma rays in the MeV –- TeV energy range, making them candidate gamma ray burst (GRB) progenitors. A number of detectors have searched for these events and reported upper limits. The upcoming High Altitude Water Cherenkov (HAWC) observatory is a next generation water Cherenkov telescope located in Mexico at 4100 m above sea level. HAWC is sensitive to gamma-rays in the energy range 50 GeV to 100 TeV. Due to its wide field-of-view and high duty-cycle, HAWC will become the prime observatory for either detecting a PBH burst or setting stringent new limits on the PBH burst rate. In this paper we present the sensitivity of HAWC to PBH bursts according to the standard model of particle physics.
}

%The keywords
\keywords{PBH, GRB, HAWC, Black Holes}

%\begin{document}
\maketitle

%Begin a section.
\section*{Introduction}

Primordial Black Holes (PBHs) are created from density inhomogeneities in many scenarios of the
early universe~\cite{Carr2010}. The initial mass of the PBH is typically roughly the horizon or Hubble mass at formation or smaller, giving possible PBH masses ranging from supermassive black hole scales down to the Planck scale. PBH production can thus have observable consequences today spanning from the very largest scales, for example influencing the development of large-scale structure in the Universe to the smallest scales, for example enhancing local dark matter clustering. PBHs, or their relics, are Cold Dark Matter candidates. For particle physics, the greatest interest is in the radiation directly emitted by the black hole. By evolving an ingoing solution past a gravitationally collapsing object, Hawking showed that a black hole will thermally emit (`evaporate') with a temperature inversely proportional to the black hole mass all available species of fundamental particles~\cite{H1}. PBHs with an initial mass of $\sim \, 5.0 \times 10^{14}$ g should be expiring now with bursts of high-energy particles~\cite{MCP}, including
gamma radiation in the MeV -- TeV energy range.

Detection of radiation from a PBH burst would provide valuable insights into the early universe and many areas of physics, as well as confirm the amalgamation of classical thermodynamics with general relativity~\cite{Carr2010}. Observations of the emitted radiation will give access to the particle physics models at energies higher than those which will ever be accessible in accelerators. Non-detection of PBHs in dedicated searches will also give important information. In the cosmological context, one of the most important motivations for PBH searches is to place limits on the spectrum of initial density fluctuations in the early universe~\cite{Carr2010}. In particular, PBHs can form from the quantum fluctuations associated with many types of inflationary scenarios~\cite{C05}. Other PBH formation mechanisms include those associated with cosmological phase transitions, topological defects or an epoch of low pressure (soft equation of state) in the early universe.
%PBHs may also be detectable by virtue of other effects. For example, PBHs with relatively large masses may be detectable by their gravitational space-time distortions in lensing or micro-lensing observations~\cite{Griest2011}. Accretion
%onto PBHs in relatively dense environments may produce distinct,
%observable emission~\cite{Trofimenko1990}. Such scenarios, however, should be
%scarce or environment-dependent and therefore difficult to use as probes of the cosmological or Galactic distributions of PBHs. On the other hand, any PBH with an initial mass of order
%$5.0 \times 10^{14}$ g should be expiring today through Hawking radiation~\cite{MCP} thereby producing a burst of gamma rays.

Evaporating PBHs are candidate gamma ray burst (GRB) progenitors.
Most GRBs are generally thought to be produced by the collapse of
massive stars (long duration GRBs) or the merger of compact
objects (short duration GRBs). However, some short duration GRBs
show behavior that may point to a different origin.
The spectra of some short duration GRBs are harder than others,
and some studies have shown that the distribution of GRBs
with durations less than 100 ms is anisotropic~\cite{CL}.
%Moreover, the offset distribution of short GRBs from their supposedly `host' galaxies is
%greater than is expected from the merger of compact objects hypothesis.
These observations may indicate a different origin for some fraction of the short GRBs.

If some short GRBs do indeed have a PBH origin, then their sources
should be located within our Galaxy and their TeV radiation
should not be attenuated by interaction with extra-galactic background photons.
Thus, we expect to see TeV gamma rays from PBH bursts.
So far, observations have not detected a TeV PBH burst.

Various direct and indirect search methods probe different distance scales when
setting PBH upper limits. Table~\ref{limit_table} gives a summary of various
search methods, distance scales that they probe and current best limits.
The upcoming High Altitude Water Cherenkov (HAWC) observatory
has the ability to either detect a PBH burst or set stringent new limits
on the PBH burst rate. In this paper we present the sensitivity of HAWC to
PBH bursts according to the standard model of Hawking radiation and particle physics.

\begin{table}[h]
\begin{center}
\begin{tabular}{|l|c|c|}
\hline Distance Scale & Limit & Method \\ \hline
Cosmological Scale & $< \, 10^{-6}$ ${\rm pc^{-3} yr^{-1}}$ & (1) \\
Galactic Scale & $<$ 0.42 ${\rm pc^{-3} yr^{-1}}$ & (2) \\
Kiloparsec Scale& $<$ 0.0012 ${\rm pc^{-3} yr^{-1}}$ & (3) \\
Parsec Scale & $<$ $4.6 \times 10^{5}$ ${\rm pc^{-3} yr^{-1}}$ & (4) \\ \hline
\end{tabular}
\caption{PBH Limits vary with distance scales: (1) from 100 MeV extragalactic $\gamma$-ray background assuming no clustering~\cite{PageHawking1976,Carr2010}, (2)
from 100 MeV anisotropy measurement~\cite{W1}, (3) from antiproton flux~\cite{Abe2012} and (4) from Very High Energy (VHE) searches~\cite{Amenomori1995}.}
\label{limit_table}
\end{center}
\end{table}

\section*{HAWC Observatory}

HAWC is a very-high-energy observatory under construction at Sierra Negra, Mexico at an altitude of 4100m. It consists of 300 water tanks with 4 photomultiplier tubes (PMT) each and will detect Cherenkov light from secondary particles created in extensive air showers induced by very-high-energy gamma rays in the energy range from ~30 GeV to 100 TeV. HAWC has two data acquisition (DAQ) systems: the main DAQ and the scaler DAQ. The main DAQ system measures the arrival direction and energy of the high-energy gamma rays by timing the arrival of particles on the ground. The direction of the original primary particle may be resolved with an error between 0.1 and 2.0 degrees depending on its energy and location in the sky. The scaler DAQ counts the number of hits in each PMT, allowing the search for excesses over the background noise. HAWC has a large field-of-view (1.8 sr or 1/7 th of the sky) and will have a high duty cycle of greater than 90\%. Thus HAWC should be able to observe high-energy emission from gamma-ray transients that extend beyond 30 GeV~\cite{Abeysekara2012}.

\section*{PBH Spectrum}

The properties of the final burst of radiation from a PBH depend on the physics
governing the production and decay of high-energy particles.
As the black hole evaporates, it loses mass and hence its
temperature and the number of distinct particle species that it emits
increase over its lifetime. In the Standard Evaporation Model (SEM)~\cite{M2, MW}, a PBH
should emit those particles whose Compton wavelengths are of the order
of the black hole size. When the black hole temperature exceeds
the Quantum Chromodynamics (QCD) confinement scale (250--300 MeV),
quarks and gluons will be directly emitted by the black hole~\cite{MW, PageHawking1976}. The quarks and gluons should fragment and hadronize as they stream away from the black hole,
analogous to the jets seen in accelerators~\cite{MW, MCP}. On astrophysical timescales, the jets
will decay into photons, neutrinos, electrons, positrons, protons and anti-protons.

Detailed studies using the SEM to simulate the particle spectra from
black holes with temperatures of $1 - 100$ GeV have shown that the
gamma-ray spectrum is dominated by the photons produced by the decay of neutral pions in the
Hawking-emitted QCD jets and is broadly peaked at photon energies of
$\sim$100 MeV. The photons which are directly Hawking-emitted and not the result of decays are visible as a much smaller peak at a much higher photon energy~\cite{MW}. As the
evaporation proceeds to higher temperatures, the greater the number of
fundamental particle degrees of freedom and the faster and more
powerful will be the final burst, with the details of the spectra
differing according to the high energy particle physics model. In this work,
we will assume the SEM as our particle physics and emission model.

\begin{figure}
\centering
\includegraphics[width=0.5\textwidth]{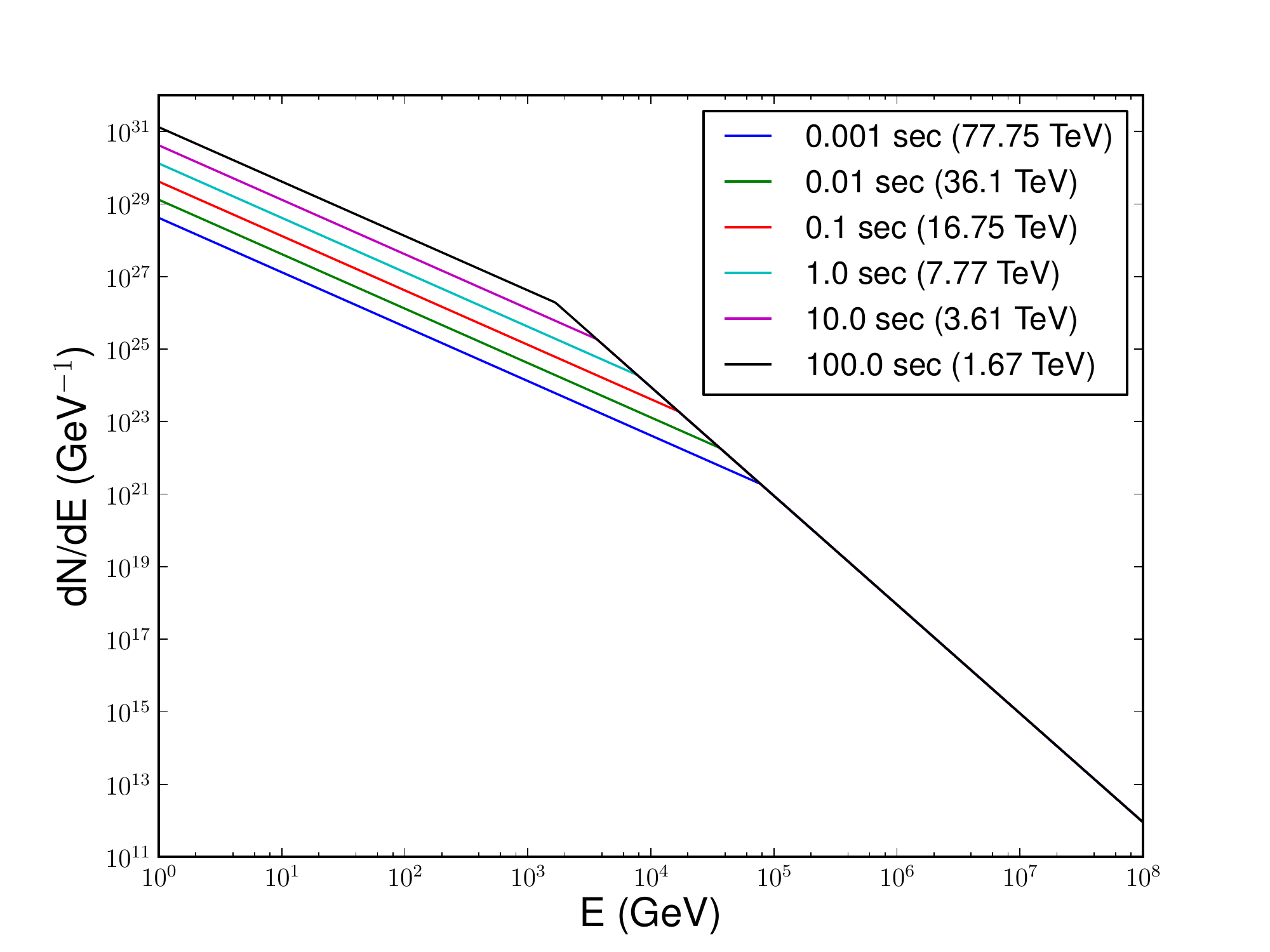}
\caption{PBH Spectrum for various remaining lifetimes. The black hole
temperature at the start of observation is also shown in parentheses.}\label{pbh_spectrum}
\end{figure}

The temperature ($T$) of a black hole depends on the remaining lifetime ($\tau$) of the black hole (the time left until the total evaporation is completed) as follows~\cite{Petkov2010}:
\begin{equation} \label{tempEq}
T = \bigg[4.7 \times 10^{11} \, \bigg(\frac{\rm{1 sec}}{\tau}\bigg) \bigg]^{1/3}\,\,\,\rm GeV.
\end{equation}
For black holes with temperatures greater than several GeVs at the start of the observation,
the time--integrated photon flux can be parameterized as (for $E > \sim$ 10 GeV)~\cite{Petkov2010}
\begin{equation} \label{photonEq}
\frac{dN}{dE} \approx 9 \times 10^{35} \begin{cases}
\big(\frac{1 GeV}{T}\big)^{3/2}\big(\frac{1 GeV}{E}\big)^{3/2},\,\,\,\,E<T \\
\big(\frac{1 GeV}{E}\big)^{3},\,\,\,\,E\ge T
\end{cases}
\end{equation}
where $E$, the gamma-ray photon energy, is measured in GeV. Figure~\ref{pbh_spectrum}
shows the PBH spectrum for various remaining lifetimes ranging from 0.001 seconds to 100 seconds.

\section*{Methodology}

\subsection*{Detectable Volume Estimation}
In order to calculate the PBH density upper limits, it is
essential to calculate the PBH detectable volume for a given detector.

In general, the expected number of photons detectable by an
observatory on the ground from a PBH burst of duration $\tau$ seconds
at a distance $r$ and zenith angle $\theta$ is
\begin{equation} \label{countsEq}
\mu(r, \theta, \tau) = \frac{(1-f)}{4 \pi r^2} \int_{E_1}^{E_2} \,\frac{dN}{dE}\, A(E,
\theta)\,dE
\end{equation}
where $f$ is the dead time of the detector, $dN/dE$ is
the gamma-ray emission spectrum integrated from remaining time $\tau$ to 0.
The values $E_1$ and $E_2$ correspond to the lower and upper bounds
of the energy range searched and $A(E,\theta)$ is the
effective area of the detector as a function of photon energy and
zenith angle. Typically the function $A(E, \theta)$ is obtained
from a simulation of the detector. For HAWC, we have parameterized the effective area for the
four zenith angle bands as $A(E) = 10^{a(\log E)^3 + b(\log E)^2 + c \log E + d} \,\,\, \rm m^2$
and the parametrization parameters are given in Table~\ref{para_table}.
\begin{table}[h]
\begin{center}
\begin{tabular}{|c|c|c|c|c|}
\hline Zenith Angle & a & b & c & d  \\ \hline
0$^{\circ}$ - 26$^{\circ}$  ($\theta_1$) & 0.1956 & -2.6778 & 12.1899 & -13.5000 \\ \hline
26$^{\circ}$ - 37$^{\circ}$ ($\theta_2$) & 0 & -0.6966 & 6.0714 & -8.2530 \\ \hline
37$^{\circ}$ - 46$^{\circ}$ ($\theta_3$) & 0 & -0.7171 & 6.5935 & -10.3238 \\ \hline
46$^{\circ}$ - 53$^{\circ}$ ($\theta_4$) & 0 & -0.5981 & 6.2712 & -11.5935 \\ \hline
\end{tabular}
\caption{Effective area parametrization parameters for various zenith angle bands.}
\label{para_table}
\end{center}
\end{table}
The minimum number of counts needed for a detection, $\mu_{\circ}(\theta_i, \tau)$, is estimated for different burst durations by finding the number
of counts required over the background for a $5 \sigma$ detection
after trials correction.

%We adopt the HAWC background rate ($R$) parametrization given
%in Ref.~\cite{Abeysekara2012} for the HAWC baseline trigger,
%\begin{equation} \label{bk_rate}
%R(\theta) = a \exp (bx + cx^2),\,\,\, {\rm where}\,\,\,x=\cos \theta - 1
%\end{equation}
%where $a$ = 0.42 Hz, $b$ = 7.0, and $c$ = -12.9. For $\theta \, <$ 25$^{\circ}$
%background stays approximately constant.

We have calculated the background rates ($R(\theta_i)$) using a Monte Carlo simulation. Using these background
rates, one can find the $\mu_{\circ}(\theta_i, \tau)$ values required for the 50\% probability of
detecting a $5\sigma$ excess after a given number of trials based on the Poisson
distribution as follows.

We define a 5$\sigma$ detection after correction for $N_t$ trials as
requiring the number of counts $n$ which would have a Poisson probability
$P$ corresponding to a corrected p-value $p_c$  given by
\begin{equation} \label{stat1}
p_c = p_o /N_t = P(\geq n|n_{\rm bk})
\end{equation}
where $p_0$ ($= 2.3 \times 10^{-7}$) is the p-value corresponding to 5$\sigma$
and $n_{\rm bk}=\tau \times R(\theta_i)$ is the number of background counts
corresponding to various burst durations.

We then find the value of $\mu_{\circ}(\theta_i, \tau)$, the amount of expected signal which
would satisfy this criterion 50\% of the time.  One expects this to be
roughly $\mu_{\circ}(\theta_i, \tau) = n - n_{bk}$; more precisely, we find $\mu_{\circ}(\tau)$
which makes the Poisson probability $P$ of finding at least $n$ counts to be
50\% according to the relation
\begin{equation} \label{stat2}
P(\geq n | n_{\rm bk} + \mu_{\circ}(\tau)) = 0.5.
\end{equation}

By substituting $\mu_{\circ}(\theta_i, \tau)$ values corresponding to various burst durations into
Equation~\ref{countsEq} and solving for $r$, we calculate the maximum distance
from which a PBH burst could be detected by the HAWC observatory for the
four zenith bands and for various burst durations,
\begin{equation} \label{distanceEq}
r_{\rm max}(\theta_i, \tau) = \sqrt{ \frac{(1-f)}{4 \pi \mu_{\circ}(\theta_i, \tau)} \int_{E_1}^{E_2} \,\frac{dN}{dE}\, A(E,
\theta_i)\,dE}.
\end{equation}
Denoting the field-of-view of the detector by
\begin{equation} \label{fov}
{\rm FOV}(\theta_{i}) = 2 \pi (1-\cos \theta_{i,\,\rm max}) {\rm sr},
\end{equation}
the detectable volume is then
\begin{equation} \label{volueEq1}
V(\tau) = \sum_{i} V(\theta_{ i}, \tau) = \frac{4}{3} \pi \sum_{i} r_{\rm max}^3(\theta_{ i}, \tau) \times \frac{{\rm effFOV}(\theta_{i})}{4\pi}
\end{equation}
where $\theta_{i}$ refers to zenith angle band and $\theta_{i,\, \rm max}$ corresponds to the maximum zenith angle in band $i$. The effFOV
is the effective field-of-view for the given zenith angle band. We calculate this by subtracting the FOV of the smaller band from the larger
band as shown below:
\begin{eqnarray} \label{volueEq2}
V(\tau) & = & \frac{1}{3} \bigg[ r_{\rm max}^3(\theta_1, \tau) \cdot {\rm FOV(\theta_1)} \nonumber \\
  &   &  + r_{\rm max}^3(\theta_2, \tau) [{\rm FOV(\theta_2)-FOV(\theta_1)}] \nonumber \\
  &   &  + r_{\rm max}^3(\theta_3, \tau) [{\rm FOV(\theta_3)-FOV(\theta_2)]} \nonumber \\
  &   &  + r_{\rm max}^3(\theta_4, \tau) [{\rm FOV(\theta_4)-FOV(\theta_3)]} \bigg]
\end{eqnarray}

\subsection*{Upper Limit Estimation}

If PBHs are uniformly distributed in the solar neighborhood, the
X\% confidence level upper limit ($UL_{X}$) to the rate density of
evaporating PBHs can be estimated as
\begin{equation}\label{ulX}
UL_{X} = \frac{m}{V \times P}
\end{equation}
if zero bursts are observed at the X\% confidence level. Here $V$ is the effective detectable volume, $P$ is the search duration and $m$ is the expected upper limit on the number of PBH evaporations given that zero bursts are observed. Note that $P_{\rm Poisson}(0|n)=1-X \rightarrow m^0 e^{-m}/0! = 1-X \rightarrow m =
- \ln(1-X) \rightarrow m = \ln (1/(1-X))$. Thus for $X=99\%$, the upper limit on the evaporating PBH rate density will be ($m=\ln 100 \approx 4.6$)

\begin{equation}\label{ul99}
UL_{99} = \frac{4.6}{V \times P}.
\end{equation}

\section*{Results}

Because we are seeking the sensitivity
in the case where there is no prior knowledge of the burst location,
we need to take into account the trials needed for the search.
If we divide the HAWC field of view, 1.8 sr, into bins of 0.7$^0$
radius, then there will be approximately 10$^4$ spatial bins
(trials) per time bin searched. The number of time bins are estimated by
dividing the total search period (5 years) by the burst duration. This results
in different numbers of trials for different burst durations. Taking these
different trials factors into account we have calculated
the $\mu_{\circ}(\theta_i, \tau)$ values corresponding to burst durations ranging from 0.001
seconds to 100 seconds for various zenith angle bands. The resulting values
for $\mu_{\circ}(\theta_i, \tau)$ are given in the Table~\ref{mu_table}.
\begin{table}[h]
\begin{center}
\begin{tabular}{|l|c|c|c|}
\hline Duration (s) & Zenith & Bk. Counts ($n_{\rm bk}$) & $\mu_{\circ}(\theta_i, \tau)$ \\ \hline
0.001 & $\theta_1$ & 0.0024 & 5.7 \\ \hline
0.001 & $\theta_2$ & 0.0012 & 4.7 \\ \hline
0.001 & $\theta_3$ & 0.0005 & 4.7 \\ \hline
0.001 & $\theta_4$ & 0.0002 & 3.7 \\ \hline
0.01 & $\theta_1$ & 0.0242 & 7.6 \\ \hline
0.01 & $\theta_2$ & 0.0115 & 6.7 \\ \hline
0.01 & $\theta_3$ & 0.0047 & 5.7 \\ \hline
0.01 & $\theta_4$ & 0.0016 & 5.7 \\ \hline
0.1 & $\theta_1$ & 0.242 & 12.4 \\ \hline
0.1 & $\theta_2$ & 0.115 & 10.6 \\ \hline
0.1 & $\theta_3$ & 0.047 & 8.6 \\ \hline
0.1 & $\theta_4$ & 0.016 & 6.7 \\ \hline
1.0 & $\theta_1$ & 2.42 & 22.2 \\ \hline
1.0 & $\theta_2$ & 1.15 & 17.5 \\ \hline
1.0 & $\theta_3$ & 0.47 & 14.2 \\ \hline
1.0 & $\theta_4$ & 0.16 & 10.5 \\ \hline
10.0 & $\theta_1$ & 24.2 & 51.5 \\ \hline
10.0 & $\theta_2$ & 11.5 & 38.2 \\ \hline
10.0 & $\theta_3$ & 4.7 & 28.0 \\ \hline
10.0 & $\theta_4$ & 1.6 & 20.1 \\ \hline
100.0 & $\theta_1$ & 242.0 & 140.7 \\ \hline
100.0 & $\theta_2$ & 115.0 & 100.7 \\ \hline
100.0 & $\theta_3$ & 47.0 & 67.7 \\ \hline
100.0 & $\theta_4$ & 16.0 & 43.7 \\ \hline
\end{tabular}
\caption{Counts needed over the background for a $5\sigma$ detection with 50\% probability
for various burst durations.}
\label{mu_table}
\end{center}
\end{table}

These $\mu_{\circ}(\theta_i, \tau)$ values are then inserted into the Equation~\ref{distanceEq} (with $E_1$=50 GeV and $E_2$=100 TeV)
and the maximum distance at which PBH burst can be seen by the HAWC observatory calculated. We have assumed negligible dead time for HAWC.
From the $r_{\rm max}$ values and Equation~\ref{volueEq2}, we calculate the effective detectable volume.
The resulting values for $r_{\rm max}$ and $V$ are shown in Table~\ref{r_table}.
\begin{table}[h]
\begin{center}
\begin{tabular}{|l|c|c|}
\hline Burst Duration (s) & $r_{\rm max}$ (pc) & Effective Volume (pc$^3$) \\ \hline
0.001 & 0.027 & 0.000010  \\ \hline
0.01 & 0.041 & 0.000030  \\ \hline
0.1 & 0.055 & 0.000070  \\ \hline
1.0 & 0.067 & 0.00012  \\ \hline
10.0 & 0.069 & 0.00011  \\ \hline
100.0 & 0.06 & 0.000070  \\ \hline
\end{tabular}
\caption{The maximum detectable distance and the detectable
effective volume for various remaining PBH lifetimes.}
\label{r_table}
\end{center}
\end{table}

Finally our 99\% confidence level upper limits
for a 5 year search period, calculated from Equation~\ref{ul99}, are shown in Figure~\ref{pbh_limits}.
%\begin{figure*}
%\centering
%\includegraphics[width=140mm]{pbh_upper_limit_plot.pdf}
%\caption{PBH Burst Rate Upper Limits from various experiments~\cite{Amenomori1995,Alexandreas,Linton,W1}.}\label{pbh_limits}
%\end{figure*}

\section*{Discussion}

\begin{figure}
\centering
\includegraphics[width=0.5\textwidth]{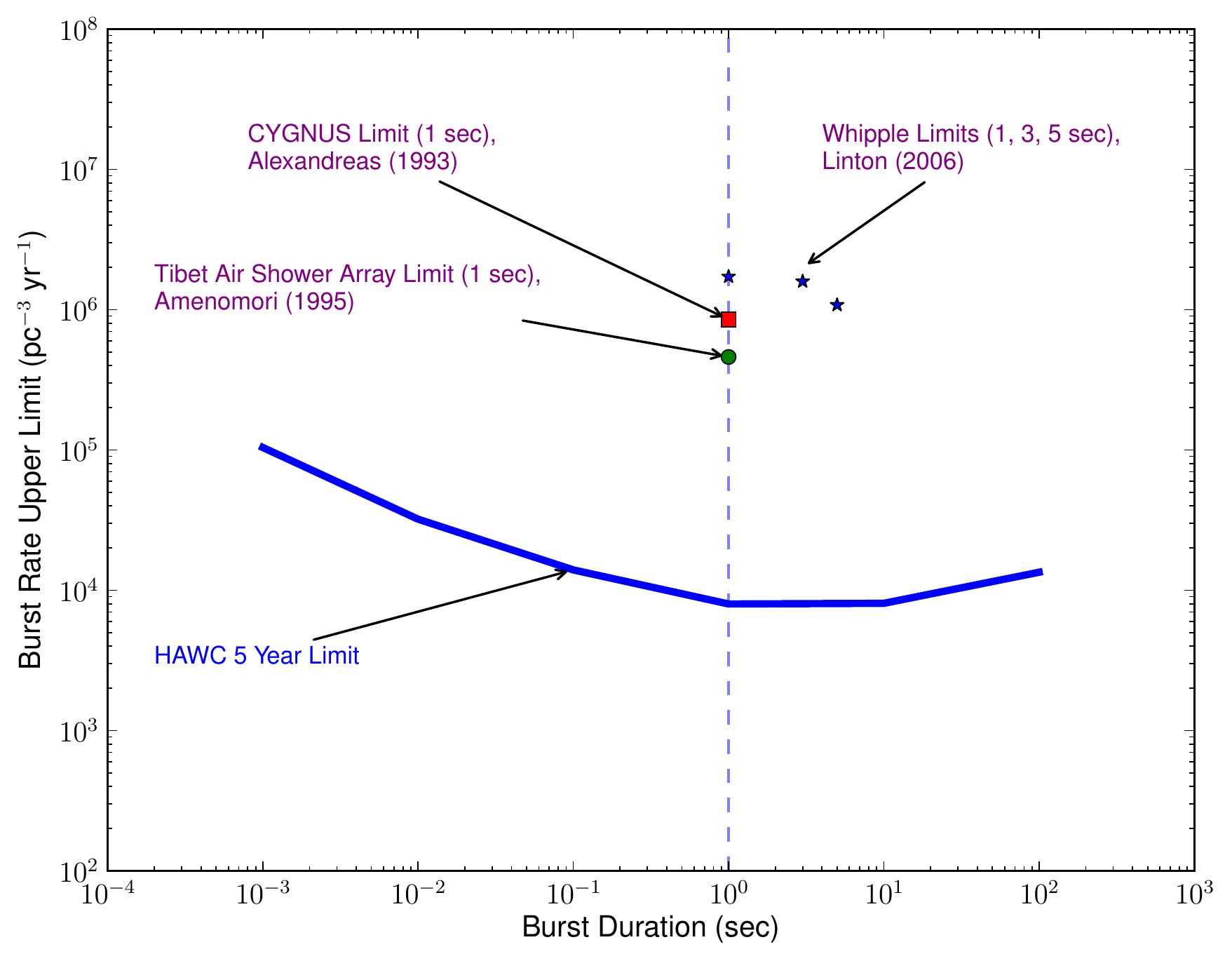}
\caption{PBH Burst Rate Upper Limits from various experiments~\cite{Amenomori1995,Alexandreas,Linton}.}\label{pbh_limits}
\end{figure}

According to Figure~\ref{pbh_limits}, a null detection with the HAWC Observatory will set upper limits which are approximately two orders of magnitude better than upper limits set by any previous PBH burst searches. The current upper limits based on null detection are also
shown in Figure~\ref{pbh_limits}~\cite{Amenomori1995,Alexandreas,Linton}.

We note that the direct search limits on the rate density of PBH bursts are weaker than
that implied by the limit on the average cosmological density of PBHs derived from the
100 MeV extragalactic gamma ray background~\cite{PageHawking1976,CM}. However, as
cold dark matter candidates, PBHs should be clustered in the Galaxy and so the local
PBH rate density should be enhanced by many orders of magnitude over the 
average cosmological PBH density. Thus a substantial number of PBHs 
that evaporate as GRBs may exist in our Galaxy. If PBHs are clustered in our 
Galactic halo, then they should also contribute an anisotropic
Galactic gamma-ray background, separable from the extragalactic background.
Wright claims that such a halo background has been detected~\cite{W1}.
The direct search limits are also weaker than that derived from the Galactic
antiproton background~\cite{Abe2012}. However the antiproton background limit depends
on the distribution of PBHs within the Galaxy and the propagation of
antiprotons through the Galaxy, as well as the production and the propagation of
the secondary antiproton component which is produced by interactions of cosmic-ray
nuclei with the interstellar gas.

The HAWC observatory has the ability to directly detect emission from nearby PBH bursts.
This capability is scientifically very important, given the large number of
early universe theories that predict PBH formation and the uncertainty in the
degree to which PBHs may cluster locally. A confirmed direct detection of an
evaporating PBH would provide unparalleled insight into high energy particle
physics and general relativity.

%\vspace*{0.5cm}
%\footnotesize{{\bf Acknowledgment:}{The ICRC 2013 is funded by FAPERJ, CNPq, FAPESP, CAPES and IUPAP.}}

\clearpage

%\end{document}

% WIMPs
\newpage
\setcounter{section}{2}
\nosection{Limits on Indirect Detection of WIMPs with the HAWC Observatory\\
{\footnotesize\sc Brian Baughman, J.~Patrick Harding}}
\setcounter{section}{0}
\setcounter{figure}{0}
\setcounter{table}{0}
\setcounter{equation}{0}
%%
% 33nd International Cosmic Ray Conference - 2013 - Rio de Janeiro, Brazil
% Template adapted from the 2011 ICRC template.
%
%\documentclass[a4paper]{article}
%
%\usepackage{icrc2013}
%%\usepackage{aas_macros}
%\usepackage{hyperref}
%\usepackage{amsmath}
%\usepackage{amssymb}

%The paper title
\title{Limits on Indirect Detection of WIMPs with the HAWC Observatory}

%The short title to appear at the header of the pages.
\shorttitle{Indirect Detection of WIMPs with HAWC}

%All paper authors
\authors{
  B.M.~Baughman$^{1}$,
  J.~Patrick Harding$^{2}$,
  for the HAWC Collaboration.
}

%All the affiliations.
\afiliations{
  $^1$ Department of Physics, University of Maryland
  College Park, MD 20742-4111
  USA
  $^2$ Physics Division
  Los Alamos National Laboratory
  Los Alamos, NM
  USA
}

%email address of the contact person
\email{bbaugh@umdgrb.umd.edu}

\newcommand{\compdate}{\ensuremath{2014}}
\newcommand{\ntanks}{\ensuremath{300}}
\newcommand{\ang}[1]{#1\ensuremath{^{\circ}}}
\newcommand{\elev}{\ensuremath{4100} m}
\newcommand{\diam}{\ensuremath{7.3} m}
\newcommand{\wdep}{\ensuremath{4.5} m}
\newcommand{\cxperad}{\ensuremath{40} m}
\newcommand{\npmts}{\ensuremath{900}}
\newcommand{\psf}[1][68]{\textrm{PSF}\ensuremath{_{#1\%}}}
\newcommand{\cx}[1][40]{\textrm{CxPE}\ensuremath{_{#1}}}
\newcommand{\nhit}{\ensuremath{N_\textrm{hit}}}
\newcommand{\trate}{\ensuremath{16} kHz}
\newcommand{\avg}[1]{\left< #1 \right>} % for average
\newcommand{\obsdays}{\ensuremath{82.8}}
\newcommand{\simobsyears}{\ensuremath{1}}
\newcommand\blfootnote[1]{%
  \begingroup
  \renewcommand\thefootnote{}\footnote{#1}%
  \addtocounter{footnote}{-1}%
  \endgroup
}

%The abstract.
\abstract{
  A growing body of evidence exists supporting the existence of Dark Matter (DM) yet its particle nature remains a mystery.
  Weakly-interacting massive particles (WIMPs) with a mass ranging from 10 GeV to 10 TeV are a popular candidate to explain the particle nature of DM.
  A WIMP hypothesis is compelling as it naturally explains the density of DM observed (thermal freeze out) and such particles are expected in some solutions to the gauge hierarchy problem.
  Such WIMPs are expected to annihilate into Standard Model particles and produce high energy photons via various processes.
  Detecting such high energy photons would provide an indirect detection of the DM particle mass.

  Below 1 TeV the Fermi-LAT provides the community with the most stringent limits of indirect detection of WIMP candidates.
  Above 1 TeV air-\v{C}erenkov telescopes such as VERITAS and H.E.S.S. have placed competitive limits using a subset of expected WIMP emission sources.
  The High Altitude Water \v{C}erenkov (HAWC) observatory complements these detectors above a few 100 GeV with its large effective area, excellent angular resolution, efficient gamma-hadron separation, and greater than 90\% duty cycle.
  HAWC is currently being deployed and operated near Puebla, Mexico.
  We present preliminary predictions for the limits for various DM annhilation channel hypotheses that we will be able to place with the HAWC Observatory.
}

%The keywords
\keywords{icrc2013, dark matter, HAWC, gamma-rays}
%

%
%\begin{document}
%

% Make title
\maketitle
\section*{Introduction}
\label{sec:intro}
Evidence for Dark Matter (DM) is present in a multitude of observations: gravitational lensing of at a multitude of scales~\cite{citeulike:10240610,citeulike:10040320,citeulike:10393545}, rotation curves of galaxies and dwarf spheroidal galaxies~\cite{citeulike:12338625,citeulike:12338628,citeulike:12338627,citeulike:12338626}, galaxy clusters~\cite{citeulike:12338630}, large-scale structure~\cite{citeulike:12338631}, and the cosmic microwave background~\cite{citeulike:10714044}; yet the particle nature of the dark matter remains a mystery~\cite{citeulike:6765814}.
Weakly-interacting massive particles (WIMPs) with a mass ranging from 10 GeV to 10 TeV are a popular candidate to explain the particle nature of DM~\cite{citeulike:6765814}.
A WIMP hypothesis is well motivated as it naturally explains the density of DM observed (thermal freeze out) and such particles are expected in some solutions to the gauge hierarchy problem.
Such WIMPs are expected to annihilate into Standard Model particles and produce high energy photons via various processes.
Detecting such high energy photons would provide an indirect detection of the DM particle mass.

Dwarf Spheroidal Galaxies are dominated by DM~\cite{citeulike:12338625,citeulike:12338588} and nearby making them ideal candidates for placing limits on $\avg{\sigma v}$ .
Furthermore, these objects have a very low flux of high-energy photons from conventional astrophysical means and are essentially background-free for the detection of annihilation signatures.
These objects have been searched for photon emission by VERITAS~\cite{citeulike:10347683}, H.E.S.S.~\cite{citeulike:12338611,citeulike:9953814}, and MAGIC~\cite{citeulike:12338546}, and Fermi Large Area Telescope (Fermi-LAT)~\cite{citeulike:9891382} and used to place limits on $\avg{\sigma v}$ vs. $M_{\chi}$ where $\chi$ denotes the DM particle.
Segue 1 has been of particular interest as it is the ``least-luminous Galaxy''
and highly DM dominated~\cite{citeulike:12338625,citeulike:12338588}.

Here we present limits on DM candidate velocity weighted cross-section vs. their masses for observations of Segue 1.
These limits use data collected on the HAWC-30 sub-detector during the deployment of HAWC and represent the most stringent limits available above 20 TeV.
Also shown are the predicted sensitivity of HAWC once completed in \compdate{}.
\section*{Instrument}
\label{sec:inst}
The High Altitude Water \v{C}erenkov (HAWC) observatory is currently being deployed near Puebla, Mexico.
When completed in \compdate{} it will consist of \ntanks{} optically--isolated galvanized steel tanks, each \diam{} in diameter and \wdep{} deep.
%~\cite{Goodman1059}.
The design is built upon the successful water \v{C}erenkov technique pioneered by Milagro~\cite{citeulike:9239358,citeulike:9239354,citeulike:9239356,citeulike:9239351,citeulike:4183441,citeulike:9240574,citeulike:9239374,citeulike:9239363}.
HAWC will have about 10 times the densely instrumented deep--water area of Milagro providing a dramatic improvement in low--energy effective area and both angular and energy resolutions.
Situated at an elevation of \elev{} (compared to 2649 m for Milagro), HAWC will detect showers which would not reach Milagro at its greater atmospheric depth, HAWC will thus have a minimum energy threshold well below that of Milagro.
The segmentation of the deep water into optically isolated tanks improves the hadron rejection efficiency over that of Milagro by a factor of about 10 at high energies.
%A detailed description of HAWC and its construction timeline can be found in these proceedings~\cite{Goodman1059}.
HAWC, with its nearly 100\% duty cycle, large field of view, and large effective area, will complement the capabilities of the Fermi-LAT~\cite{citeulike:4118358,citeulike:5758938,citeulike:7058605,citeulike:9339060,citeulike:6657466} and make it an excellent observatory for transient objects.

HAWC's modularity allows data taking during deployment; as tanks are instrumented and verified they are placed into the data stream.
The HAWC-30 sub-detector consists of approximately 30 tanks in a rough equilateral triangle.
Data presented herein use the HAWC-30 sub-detector.
\section*{Method}
\label{sec:method}
We use the standard method of predicting the gamma-ray flux due to various annihilation hypothesis and source morphologies~\cite{citeulike:12338548}.
This method has been employed in searches conducted by the VERITAS~\cite{citeulike:10347683}, H.E.S.S.~\cite{citeulike:12338611,citeulike:9953814}, and MAGIC~\cite{citeulike:12338546} Imaging Atmospheric \v{C}erenkov Telescopes (IACTs), and by the Fermi Large Area Telescope (Fermi-LAT)~\cite{citeulike:9891382}.
A more detailed description of the full method can be found in Abazajian and Harding~\cite{citeulike:9953814}
Here we examine four annihilation channels:
\begin{itemize}
  \item $ \chi \bar{\chi} \rightarrow W \bar{W} $
%  \item $ \chi \bar{\chi} \rightarrow \mu \bar{\mu} $
  \item $ \chi \bar{\chi} \rightarrow t \bar{t} $
  \item $ \chi \bar{\chi} \rightarrow b \bar{b} $
  \item $ \chi \bar{\chi} \rightarrow \tau \bar{\tau} $
\end{itemize}

\begin{figure}[t]
  \begin{center}
    \includegraphics[width=0.45\textwidth]{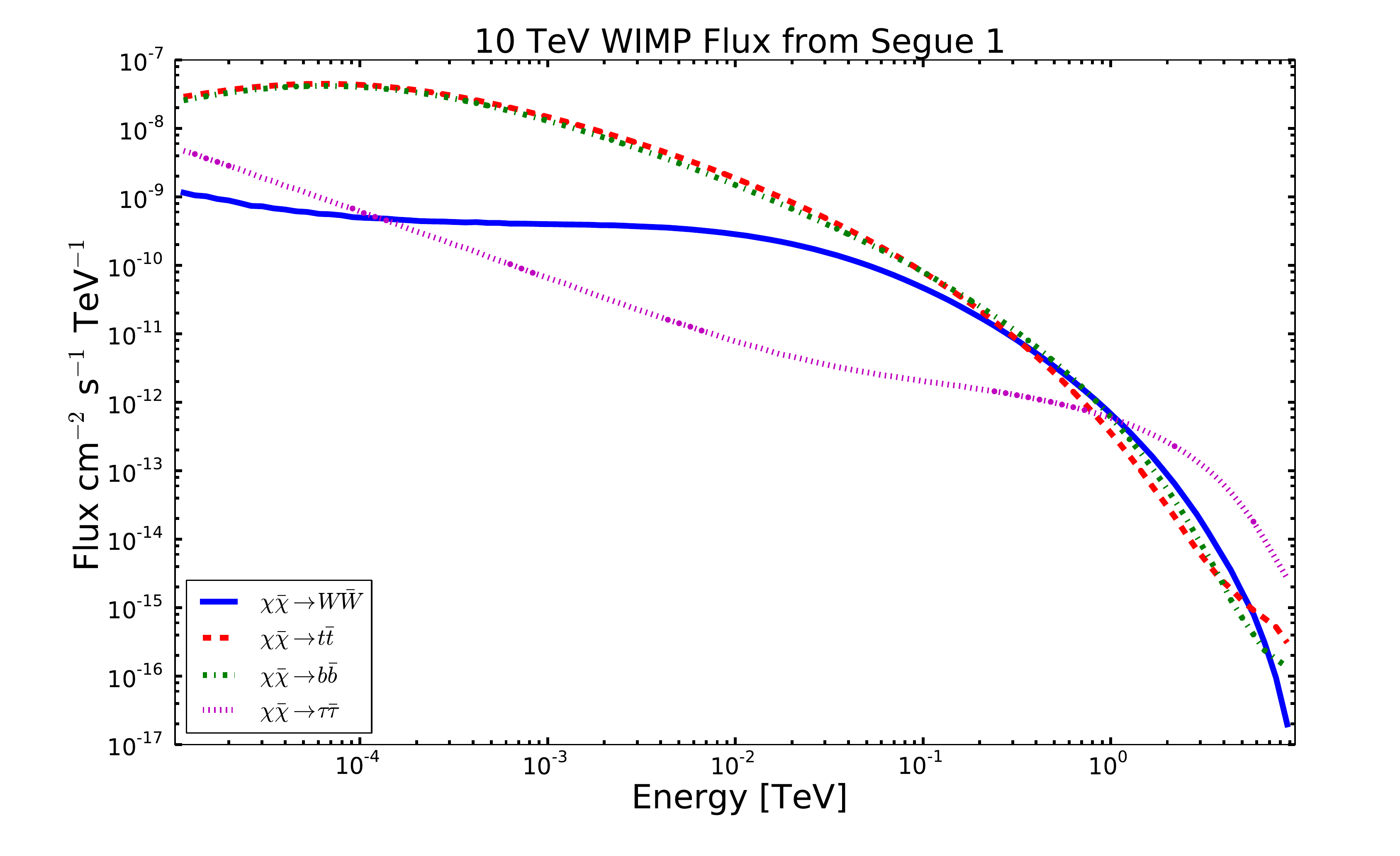}
    \caption{
      Predicted differential photon flux, given by Eqn. \ref{eqn:wimpflux}, from Segue 1 for 10 TeV WIMP candidates for various annihilation channels.
    }
    \label{fig:diffflux}
  \end{center}
\end{figure}

The differential photon flux for any annihilation channel is written as:
\begin{equation}
  \label{eqn:wimpflux}
  \frac{d^3 N_{\gamma}}{dE dt dA} = \frac{1}{4\pi} \frac{\avg{\sigma v}}{2 M_{\chi}^2} \frac{dN_\gamma}{dE}  J(\psi)
\end{equation}
where $\avg{\sigma v}$ is the velocity-weighted average cross section, $M_{\chi}$ is the mass of the presumed dark matter, ${dN_{\gamma} / dE}$ is the gamma-ray energy distribution per annihilation from the subsequent showering and decay of the annihilation products.
The photon flux observable at Earth is determined using PYTHIA ~\cite{citeulike:620804} by setting the center-of-mass energy to twice $M_{\chi}$ and simulating the shower for each annihilation channel.
An example of ${d^3\Phi_{\gamma}}/{dE dt dA}$ is shown in Figure \ref{fig:diffflux} for a WIMP mass of 10 TeV.
The shapes of these spectra generally scale with WIMP mass, e.g. a 1 TeV WIMP annihilating into $\tau\bar{\tau}$ would have a local maxima at around 300 GeV instead of 3 TeV shown in the figure.

The $ J ( \psi ) $ is the integral of the squared DM density ($\rho$) over the line of sight ($\psi$) and field of view ($d\Omega$) as shown in Eqn. \ref{eqn:astrofac}.
\begin{equation}
  \label{eqn:astrofac}
  J(\psi) = \int_{\Delta \Omega} \int_{l.o.s.}\rho_{\chi}^2(l(\psi)) dl d\Omega
\end{equation}
This factor contains all the astrophysical information about the source being examined and is generally constrained by other observations of the object being studied.
As the gamma-ray flux is proportional to this factor, sources with a large $ J ( \psi ) $ and minimal gamma-ray flux due to non-DM sources are ideal.
The $ J ( \psi ) $ factor for Segue 1 is $7.7 \cdot 10^{18}$ GeV$^{-2}$ cm$^{-5}$ sr.

Using the above formalism for each source, annihilation channel, and WIMP mass, limits on $\avg{\sigma v}$ can be placed.
The limiting value of $\avg{\sigma v}$ is found by finding the maximum value consistent with our data (or simulation) at the two $\sigma$ level (approximately 95\%).
HAWC is sensitive to DM masses of about 1 TeV and above.
\section*{Results}
\label{sec:results}
\begin{figure}[t]
  \begin{center}
    \includegraphics[width=0.45\textwidth]{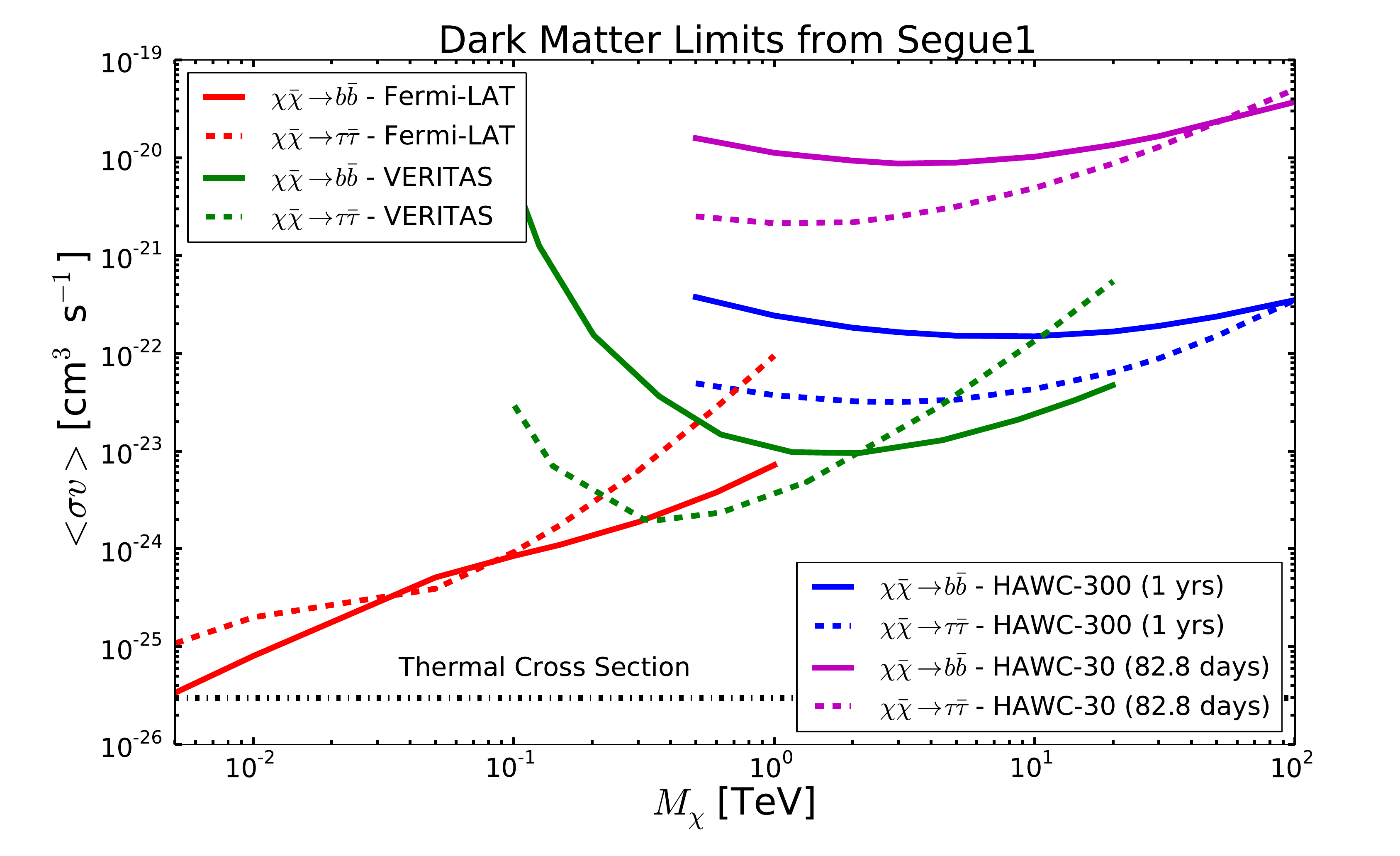}
    \caption{
      Current limits on $\avg{\sigma v}$ vs. $M_{\chi}$ in the annihilation channels $ \chi \bar{\chi} \rightarrow \tau \bar{\tau} $ and  $ \chi \bar{\chi} \rightarrow b \bar{b} $ from observations of Segue 1 made by Fermi-LAT (24 months)~\cite{citeulike:9891382} and VERITAS (50 hrs)~\cite{citeulike:10347683}.
      Preliminary HAWC-30 sensitivity for \obsdays{} day of observation and anticipated sensitivity of HAWC-300 for \simobsyears{} years of observation.
    }
    \label{fig:segue1-comp}
  \end{center}
\end{figure}

\begin{figure}[t]
  \begin{center}
    \includegraphics[width=0.45\textwidth]{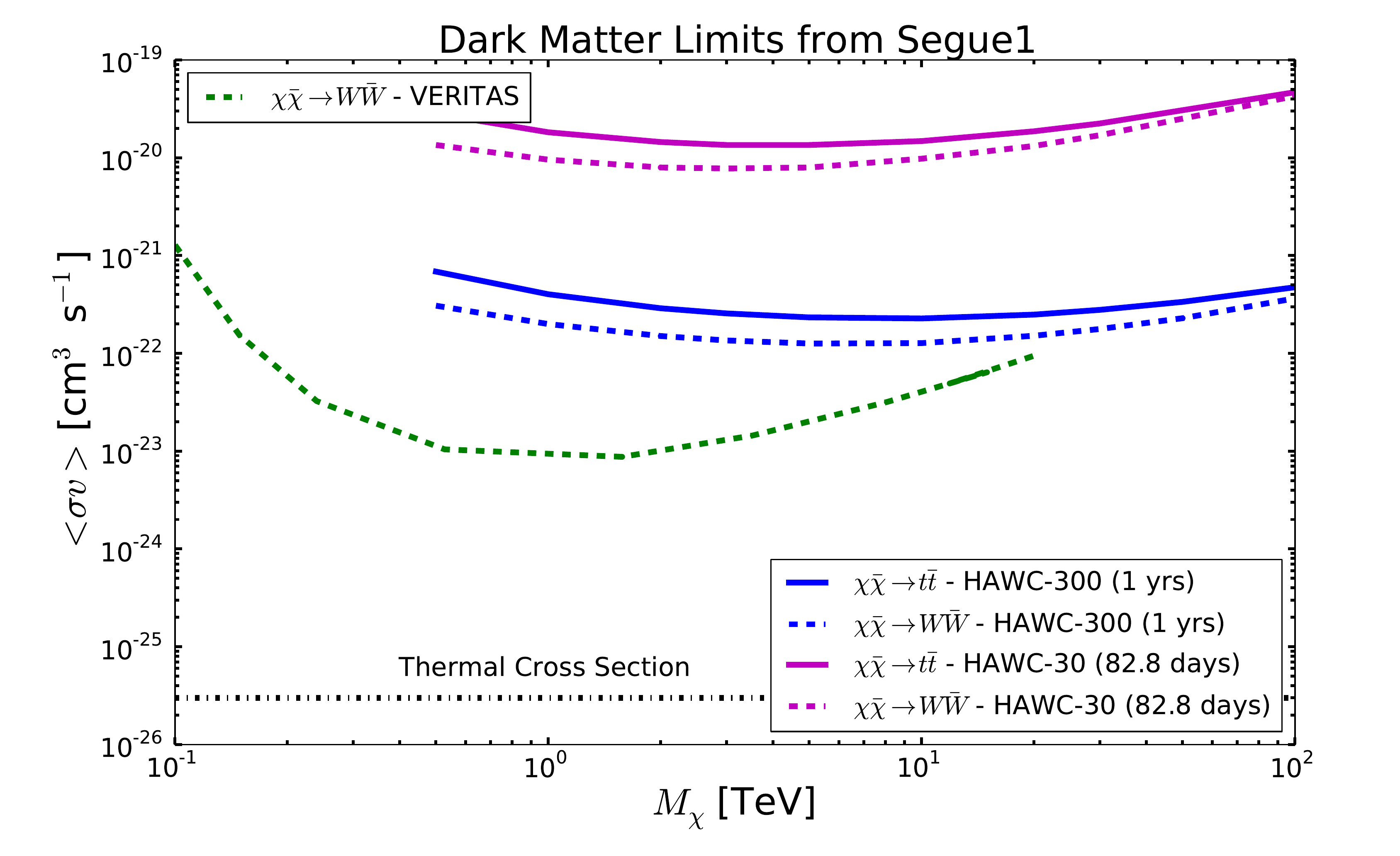}
    \caption{
      Preliminiary limits on $\avg{\sigma v}$ vs. $M_{\chi}$ in the annihilation channels $ \chi \bar{\chi} \rightarrow t \bar{t} $ and  $ \chi \bar{\chi} \rightarrow W \bar{W} $ from Segue 1.
    }
    \label{fig:segue1-remain}
  \end{center}
\end{figure}

Figures \ref{fig:segue1-comp} and \ref{fig:segue1-remain} show the limits obtained from observations of Segue 1 with Fermi-LAT ( 24 months), VERITAS (50 hrs), and HAWC-30 (\obsdays{} days) as well as the anticipated limits from HAWC-300 (\simobsyears{} years).
The cross section needed to produce all of the dark matter thermally in the early universe is shown for comparison.
The cross section at z=0 could be substantially higher (10$^3$ or 10$^4$) than during the thermal freeze out if the mediators of the annihilation are massive bosons, the so-called Sommerfeld enhancement effect (e.g.~\cite{citeulike:12338601}).
Above 1 TeV this possibility is very real:
the W and Z bosons are already-known particles that must enhance the cross section at low velocities if they participate in the annihilation.
With such cross-section boosts and potential boosts (perhaps as much as 10x) from un-modeled dark matter substructure, thermally-produced WIMPs may be accessible with HAWC.
\section*{Conclusion}
\label{sec:conclusion}
HAWC-30, a small fraction of the final area of HAWC-300, is already providing limits from indirect detection on $\avg{\sigma v}$ vs. $M_{\chi}$ as seen in Figs. \ref{fig:segue1-comp} and \ref{fig:segue1-remain}.
As HAWC continues to be deployed its sensitivity will improve from increases in effective area, angular resolution, and gamma/hadron separation.
Observations with HAWC on DM dominated objects such as Segue 1 will provide competitive limits on dark matter with masses above 1 TeV.
Additionally, much like the Fermi-LAT, HAWC can be used to observe every dwarf spheroidal galaxy in its field of view (even ones that are newly discovered) as well as to stack the signals from many sources much like analyses of Fermi-LAT data~\cite{citeulike:9891382}.

\section*{Acknowledgments}

We acknowledge the support from: US National Science Foundation (NSF); US
Department of Energy Office of High-Energy Physics; The Laboratory Directed
Research and Development (LDRD) program of Los Alamos National Laboratory;
Consejo Nacional de Ciencia y Tecnolog\'{\i}a (CONACyT), M\'exico; Red de
F\'{\i}sica de Altas Energ\'{\i}as, M\'exico; DGAPA-UNAM, M\'exico; and the
University of Wisconsin Alumni Research Foundation.

% Bibliography

\clearpage
%
%\end{document}

% LIV
\newpage
\setcounter{section}{3}
\nosection{Sensitivity of the HAWC Detector to Violations of Lorentz
Invariance\\
{\footnotesize\sc Lukas Nellen}}
\setcounter{section}{0}
\setcounter{figure}{0}
\setcounter{table}{0}
\setcounter{equation}{0}
%%
% 33rd International Cosmic Ray Conference - 2013 - Rio de Janeiro, Brazil
% Template adapted from the 2011 ICRC template.
%
%\documentclass[a4paper]{article}
%
%\usepackage{icrc2013}
%\usepackage[english]{babel}
%\usepackage[utf8]{inputenc}
%\usepackage[T1]{fontenc}
%%\usepackage{showkeys}
%\usepackage{cite}
\newcommand{\U}[1]{\,\mathrm{#1}}
\newcommand{\UU}[1]{\mathrm{#1}}
\newcommand{\EQG}{E_\mathrm{QG}}

%The paper title
\title{Sensitivity of the HAWC Detector to Violations of Lorentz Invariance}

%The short title to appear at the header of the pages.
\shorttitle{Sensitivity of the HAWC Detector to Violations of Lorentz Invariance}

%All paper authors
\authors{
Lukas Nellen$^{1}$
for the HAWC Collaboration.
}

%All the affiliations.
\afiliations{
$^1$ I de Ciencias Nucleares, UNAM, Mexico 
}

%email address of the contact person
\email{lukas@nucleares.unam.mx}

%The abstract.
\abstract{Lorentz invariance is believed to be a fundamental symmetry
  of the universe. Many theories of quantum gravity, however, break
  Lorentz invariance at small scales and high energies explicitly. It
  is, therefore, of great interest to be able to place limits on some
  model and, if possible, even to observe this effect. The observation
  of a violation of Lorentz invariance would revolutionize our view of
  the universe and probe physics at energy scales not attainable with
  earthbound accelerators. Gamma-ray bursts provide an ideal
  laboratory to search for such phenomena. The combination of extreme
  distance (billions of light years), high energy emission (up to at
  least $30 \U{GeV}$), and short duration (burst durations of less than
  one-second have been observed), allows one to measure the relative
  speed of different energy photons to a part in $~10^{16}$.  In this
  paper we will discuss current limits and the prospect for HAWC to
  improve upon these limits.
}

%The keywords
\keywords{HAWC, gamma rays, Lorentz Invariance Violation}

%\begin{document}
\maketitle

\section*{Introduction}
\label{sec:introduction}

In this paper, we evaluate the potential of the HAWC observatory
\cite{bib:hawctalk, bib:hawcweb} to
place limits on theoretical models of some classes of theories, which
predict the violations of Lorentz Invariance.

Lorentz invariance is a fundamental symmetry in modern physics, in
particular of quantum field theories (QFTs) \cite{bib:peskin}. 
Many theories and models developed to unify QFTs with General
Relativity predict the violation of Lorentz Invariance at very small
scale or very high energies \cite{bib:smolin}, which can lead to
observable effects \cite{bib:mattingly}. 

Based on dimensional arguments, it is typically expected that effects
of quantum gravity (QG) and therefore Lorentz Invariance Violation
(LIV) occur at the Planck scale, characterised by the Planck Mass
$m_\mathrm{pl} = \sqrt{\hbar c / G} \approx 10^{19}\U{GeV}$. The
Planck Scale is out of direct reach of any current or foreseeable
accelerator. Nevertheless, it is possible that effects of QG 
lead to small but observable effects at lower, accesible, energies. Of
particular interest in gamma ray observations are models which predict
a modification of the dispersion relation for photons, which implies
deviation of the velocity of photons from the speed of light in vacuum
$c$. The leading term of the modified dispersion relation is 
\cite{bib:bolmont2011}
\begin{equation}
E^2 \simeq p^2c^2 \Biggl(1 + 
  \xi_n \biggl(\frac{pc}{E_\mathrm{QG}}\biggr)^n\Biggr).
\label{eq:dispersion}
\end{equation}
In this equation, $E_\mathrm{QG}$ is the energy scale where QG effects
set in and $\xi_n$ an expansion coefficient. The order of the leading
term is model dependent. Models typically considered have linear or
quadratic leading terms.

In studies using gamma
ray sources, it is common to combine the Planck scale and the leading
coefficient into one
scale variable $E_\mathrm{QG}^{(n)} = E_\mathrm{QG} \xi_n^{-1/n}$ and
to quote limits for this variable.
The modified dispersion relation leads to a modification in the
propagation speed of photons
\begin{equation}
  \label{eq:speed}
  v = \frac{\mathrm{d}E}{\mathrm{d}p} 
  \approx c\Biggl( 1 +
  \frac{n+1}{2} \biggl(\frac{pc}{E_\mathrm{QG}^{(n)}}\biggr)^n \Biggr).
\end{equation}

The energy dependence of the propagation speed of photons translates
into a difference in the arrival time $\Delta t$ of photons emitted at
the same time. For near-by sources, like pulsars, the relation between
the distance of the source and the propagation time is straightforward
and leads to the following relation
\begin{equation}
  \label{eq:dt-near}
  \Delta t = \frac{(n+1)d}{2c} 
             \frac{\Delta E^n}{\bigl(E_\mathrm{QG}^{(n)}\bigr)^n}
           \approx \frac{(n+1)d}{2c} 
             \frac{E_\mathrm{max}^n}{\bigl(E_\mathrm{QG}^{(n)}\bigr)^n}
\end{equation}
where $\Delta E^n = E_\mathrm{max}^n - E_\mathrm{min}^n$, which can be
approximated by the the maximum photon energy as $\Delta E^n \approx
E_\mathrm{max}^n$. The distance to the source is $d$.

For sources at cosmological distances, like active galactic nuclei
(AGN) or gamma ray bursts (GRB), one has to
include the effects of the redshift during propagation and the
non-trivial metric of the expanding universe to obtain \cite{bib:bolmont2008}
\begin{equation}
  \label{eq:dt-far}
  \Delta t = \frac{n+1}{2} H_0^{-1} \frac{E_\mathrm{max}^n}{\bigl(E_\mathrm{QG}^{(n)}\bigr)^n}
\int_0^z \frac{(1+z')^n}{h(z')}\,\mathrm{d}z'
\end{equation}
with $h(z) = \sqrt{\Omega_\Lambda + \Omega_m (1+z)^3}$ and using
the cosmological parameters $H_0 = 70 \U{km} \U{s}^{-1} \U{Mpc}^{-1}$,
$\Omega_\Lambda = 0.7$, and $\Omega_m = 0.3$.

\section*{LIV limits}
\label{sec:liv-limits}

\begin{table*}[htb]
  \centering
  \begin{tabular}{lccccccc}
    \hline\hline
    \multicolumn{1}{c}{Source}& Experiment & Limit on $\EQG^{(1)}$ &
    Limit on $\EQG^{(2)}$ & Distance & $\Delta t$ & $E_\mathrm{max}$ &
    Ref.\\
    \hline \\[-2.1ex]
    Crab   & VERITAS & $3 \cdot 10^{17} \U{GeV}$ & $7 \cdot 10^9 \U{GeV}$ &
      $2.2 \U{kpc}$ & $100 \U{\mu s}$ & $120 \U{GeV}$ &
      \cite{bib:otte2011}\\
    GRB090510 & Fermi/LAT & $1.5 \cdot 10^{19} \U{GeV}$ & 
      $3 \cdot 10^{10} \U{GeV}$ &
      $z=0.903$ & $829 \U{ms}$ & $31 \U{GeV}$ & 
      \cite{bib:abdo2009a}\\
    PKS 2155-304 & HESS & $2.1 \cdot 10^{18} \U{GeV}$ &
      $6.4 \cdot 10^{10} \U{GeV}$ &
      $z = 0.116$ & \multicolumn{2}{c}{likelihood fit} &
      \cite{bib:abramowski2011}\\
    \hline \\[-2.1ex]
    HAWC Pulsar ref. & HAWC & $10^{17} \U{GeV}$ & 
      $9 \cdot 10^9 \U{GeV}$ & $2 \U{kpc}$ & $1 \U{ms}$ & 
      $500 \U{GeV}$ &\\
    HAWC GRB ref. & HAWC & $4.9 \cdot 10^{19}\U{GeV}$ &
      $1.5 \cdot 10^{11}\U{GeV}$ & $z = 1$ & $1 \U{s}$ &
      $100 \U{GeV}$ & \\
    \hline\hline
  \end{tabular}
  \caption{Compilation of the most stringent results on LIV published
    and the 
    potential of the HAWC observatory, based on the reference
    scenarios described in section~\ref{sec:hawc-potential-liv}.}
  \label{tab:limits}
\end{table*}

Observations of Pulsars, GRBs, and AGN have been used to establish
limits on the scale of QG. For a compilation of results, see
\cite{bib:bolmont2011} and references therein. Some additional results are in
\cite{bib:otte2011,bib:Amelino-Camelia2013}. The most difficult part in
establishing stringent limits is estimating the possible propagation
delay. The simplest assumption to make is the simultaneous emission of
all photons, Using this assumptions means one over-estimates the time
difference $\Delta t$, since astrophysical effects in the sources will
themselves spread out the emission of photons over an extended period
of time. Precise modelling and sophisticated analysis techniques can
help to improve limits on $\Delta t$ and thereby improve the derived limits.

The best limits from pulsars have been reported by VERITAS
\cite{bib:otte2011} to be $\EQG^{(1)} > 3\cdot 10^{17}\U{GeV}$ and 
$\EQG^{(2)} > 7 \cdot 10^9 \U{GeV}$.

The most stringent limits from GRBs have been established by the
Fermi/LAT collaboration, based on the observation of GRB090510
\cite{bib:abdo2009a} to be $\EQG^{(1)} > 1.5 \cdot 10^{19}\U{GeV}$ and 
$\EQG^{(2)} > 3 \cdot 10^{10}\U{GeV}$. Somewhat less stringent limits
have been set using data from GRB080916C \cite{bib:abdo2009b}.

The limit on the quadratic term has been improved by HESS using the
observation of the flare of PKS 2155-304 on MJD 53944 
\cite{bib:abramowski2011} to $\EQG^{(2)} > 6.4 \cdot 10^{10}\U{GeV}$.

\section*{HAWC potential for LIV limits}
\label{sec:hawc-potential-liv}

Motivated by the sources used in the analyses quoted in
section~\ref{sec:liv-limits}, we establish reference scenarios for
GRB and pulsar observation to establish the potential of HAWC for
setting limits on LIV models.

Our reference scenario for pulsars is motivated by the VERITAS
observation of the 
Crab pulsar up to $120\U{GeV}$ \cite{bib:aliu2011} and the limits of
LIV derived from that observation \cite{bib:otte2011}. We assume a
limit $\Delta t = 1\U{ms}$, a factor of 10 worse than the VERITAS
limit. This gets compensated by assuming a pulsar with $E_\mathrm{max}
= 500\U{GeV}$ at a distance of $2 \U{kpc}$.
Based on this, we can establish limits for the linear term $\EQG^{(1)}$ up to
$10^{17}\U{GeV}$ and for the quadratic term $\EQG^{(2)}$ of up to 
%\linebreak{}
$9 \cdot 10^9 \U{GeV}$. The assumption of an increase in
the observed $E_\mathrm{max}$ allows for an improved limit on the
quadratic term, despite the increase in the assumed limit on the time
difference. One has to assume that such a source, if it exists, will
also be observed by IACTs. It will remain to be seen if the HAWC's
ability to monitor a pulsar continuously will allow us to reach a
competitive limits on differences in propagation time, needed to
establish stringent limits. 

To establish a reference scenario for LIV limits from GRBs, we look at
the properties of the short burst GRB090510 \cite{bib:abdo2009a} and
of the long burst GRB130427A \cite{bib:Amelino-Camelia2013}. As a
reference, we select
a short burst with $\Delta t = 1 \U{sec}$ at a redshift of $z = 1$
with a maximum observed photon energy of $E_\mathrm{max} = 100
\U{GeV}$, a burst which is detectable if it occurs in the field of
view of HAWC \cite{bib:hawcgrb}.
The time difference and redshift of the reference scenario are clearly
compatible with values of 
GRB0800916C \cite{bib:abdo2009a}, GRB090510 \cite{bib:abdo2009b}, and
GRB130427A \cite{bib:Amelino-Camelia2013}. The maximum photon energy
at the source in our reference scenario is $200 \U{GeV}$. We believe
this to be slightly optimistic, but justified, given existing
observations. The
maximum photon energies reported for the GRBs mentioned above are
$13\U{GeV}$, $31 \U{GeV}$, and $94 \U{GeV}$, which translate to 
$69 \U{GeV}$, $55 \U{GeV}$, and $125 \U{GeV}$.
In this scenario, a limit of $4.9 \cdot 10^{19}\U{GeV}$ for
the linear term $\EQG^{(1)}$ and $1.5 \cdot 10^{11}\U{GeV}$ for
$\EQG^{(2)}$. Comparing these number with the limits reported in
section~\ref{sec:liv-limits} shows that, in this scenario, HAWC has
the potential to improve on the currently established limits both for
the linear and quadratic terms.

It is unlikely that HAWC will be able to improve limits based on
observations of AGN flares by IACTs. The lower statistics of HAWC
implies less detail in light curves, and in turn a less restrictive
estimate of $\Delta t$.

Of the sources considered here, pulsars have the advantage that they
are long-lived and therefore reliably detectable. HAWC's ability to
monitor a source continuously competes with higher statistics, but
time limited, observations by IACTs. One-off transients like GRBs have
the largest potential for setting stringent limit. HAWC fares well in
this scenario, since it has a higher energy reach than satellite
instruments, and a larger field of view than IACTs, which increases
the chance of detecting short transients. The duration of AGN flares
is sufficiently large for IACTs to slew into position to observe. It
is unlikely that HAWC will be able to compete in this area, assuming
similar observation times.

% Whereas the short timescale from
% seconds to a few minutes of GRBs makes it hard to slew an ACT to the 

% , since the time-scales of such
% flares allow for detailed observations of such events, which will 

\section*{Conclusions}

We demonstrated the potential of the HAWC gamma ray observatory to
study possible deviations of the speed of photons from the speed of
light $c$, which are predicted by some classes of models with LIV. We
conclude that HAWC is capable of setting competitive limits on the
scale of LIV in such models, especially using GRB
observations. 

\section*{Acknowledgments}

We acknowledge the support from: US National Science Foundation (NSF); US
Department of Energy Office of High-Energy Physics; The Laboratory Directed
Research and Development (LDRD) program of Los Alamos National Laboratory;
Consejo Nacional de Ciencia y Tecnología (CONACyT), México; Red de Física de
Altas Energias, Mexico; DGAPA-UNAM (PAPIIT-IN113612), México; VIEP-BUAP; the
University of Wisconsin Alumni Research Foundation; the Institute of
Geophysics; and Planetary Physics at Los Alamos National Lab.

%Also it is possible to include the Acknowledgment as a footnote.}}%The ICRC 2013 is funded by FAPERJ, CNPq, FAPESP, CAPES and IUPAP.}}

\clearpage

%\end{document}

% IGMF
\newpage
\setcounter{section}{4}
\nosection{HAWC Contributions to IGMF Studies\\
{\footnotesize\sc Thomas Weisgarber}}
\setcounter{section}{0}
\setcounter{figure}{0}
\setcounter{table}{0}
\setcounter{equation}{0}
%%
% 33nd International Cosmic Ray Conference - 2013 - Rio de Janeiro, Brazil
% Paper based on the 2013 ICRC template.
%
%\documentclass[a4paper]{article}
%
%\usepackage{icrc2013}
%\usepackage[english]{babel}

%The paper title
\title{HAWC Contributions to IGMF Studies}

%The short title to appear at the header of the pages.
\shorttitle{HAWC Contributions to IGMF Studies}

%All paper authors
\authors{
Thomas Weisgarber$^{1}$ for the HAWC Collaboration.
}

%All the affiliations.
\afiliations{
$^1$ Department of Physics, University of Wisconsin\textemdash Madison, Madison, WI USA \\
}

%email address of the contact person
\email{weisgarber@physics.wisc.edu}

%The abstract.
\abstract{
The intergalactic magnetic field (IGMF), presumed to exist in the void regions between galaxy clusters, may play a role in galaxy formation, contain information about conditions in the early universe, or influence the trajectories of cosmic rays of extragalactic origin.
Recent studies have attempted to measure the IGMF by searching for its influence on the pair cascades produced when very high energy gamma rays from blazars interact with the extragalactic background light (EBL).
The analysis of simultaneous data from imaging atmospheric Cherenkov telescopes (IACTs) and the Fermi Gamma Ray Space Telescope (Fermi) suggests that the strength of the IGMF may be greater than $10^{-15}$ gauss.
However, this conclusion relies on assumptions about the properties of the source that are difficult to verify with existing IACT observations.
The High Altitude Water Cherenkov (HAWC) Observatory, currently under construction on the slopes of Sierra Negra in Mexico, is uniquely situated to contribute to measurements of the IGMF.
With an instantaneous field of view of 2 sr, sensitivity to gamma rays with energies above 50 GeV, and a large duty cycle, HAWC will provide unbiased observations of the average fluxes from blazars and accurate measurements of the attenuation of gamma rays due to interactions with the EBL.
In this work, we present the capability of HAWC to contribute to measurements of the IGMF via observations of the delayed secondary flux following a bright blazar flare.
}

%The keywords
\keywords{HAWC, blazars, gamma rays, IGMF.}

%\begin{document}
\maketitle

%Begin a section.
\section*{The Intergalactic Magnetic Field}

Magnetic fields are known to exist ubiquitously within the galaxies, filaments, and clusters that comprise the large-scale structure of the universe.
However, conclusive evidence for the existence of a magnetic field in the void regions that dominate the volume of the universe remains elusive.
The voids can become magnetized during phase transitions in the early universe, after which components of the field with correlation lengths large enough to withstand magnetic diffusion over a Hubble time evolve in strength only due to the cosmic expansion~\cite{bib:grasso}.
Referred to as the intergalactic magnetic field (IGMF), the field in the voids may provide the seeds for magnetohydrodynamic processes that generate the fields presently observed in galaxies and clusters.
Seed fields with present-day values as small as $B=10^{-20}$ gauss could readily explain the observed galactic and cluster fields when adiabatic compression and dynamo generation are taken into account~\cite{bib:widrow}.
At present, however, Faraday rotation measurements of distant quasars constrain the IGMF strength to be smaller than $10^{-9}$ gauss~\cite{bib:neronov1}, leaving a wide range of field strengths unexplored and the seed-field hypothesis largely untested.

In recent years, a new technique has been developed that employs gamma-ray observations of distant blazars to measure the intergalactic magnetic field (IGMF) in the unexplored range from $10^{-18}$ to $10^{-14}$ gauss~\cite{bib:neronov1,bib:eungwanichayapant,bib:dolag1,bib:elyiv}.
Gamma rays with energies above a few hundred GeV interact with the extragalactic background light (EBL), producing electron-positron pairs whose trajectories are sensitive to the strength of the IGMF.
The pairs scatter target photons, primarily from the cosmic microwave background (CMB), via the inverse Compton process, producing a secondary cascade of gamma rays.
Owing to the deflection of the pairs, the influence of the IGMF manifests in the time profile~\cite{bib:plaga,bib:ichiki,bib:murase} and angular extent~\cite{bib:aharonian,bib:ahlers} of the cascade emission.

Several recent studies~\cite{bib:neronov2,bib:tavecchio1,bib:tavecchio2,bib:dermer,bib:essey,bib:huan,bib:dolag2,bib:taylor} have used ground-based imaging atmospheric Cherenkov telescopes (IACTs) to measure the attenuated direct flux which gives rise to the cascade.
These studies have found that, due to the absence of any observed cascade by the Fermi Gamma Ray Space Telescope (Fermi), the IGMF strength is likely larger than $10^{-15}$ gauss.
However, the conclusions of these studies rest heavily on the assumption that the sparsely sampled measurements of the flux from a few sources are representative of the flux from those sources over a period of several years, and a change of 50\% in the average flux could invalidate their conclusions~\cite{bib:arlen}.
Unbiased measurements of the average flux from the sources are therefore required to interpret the results of these non-observations of the cascade by Fermi.

\section*{The HAWC Dectector}

The High Altitude Water Cherenkov (HAWC) Observatory, currently under construction on the slopes of Sierra Negra in the Mexican state of Puebla, is uniquely situated to contribute to studies of the IGMF.
The characteristics and general capabilities of HAWC are described in detail in~\cite{bib:hawc}.
Boasting a large field of view of 2 sr and a duty cycle above 90\%, HAWC will observe all blazars with declinations between $-20^\circ$ and $60^\circ$ in an unbiased manner every day.
Among many other science goals, HAWC will monitor the long-term properties of the gamma-ray flux from these blazars, and it also will search for strong flares.
In this work, we focus on the capability of HAWC to detect the IGMF based on the timing information available from a bright blazar flare.

\section*{Blazar Flares}

Blazars are known to be highly variable sources, often flaring to fluxes more than an order of magnitude higher than their quiescent emission for time periods ranging from minutes to years.
For example, Markarian 421 (Mrk421), the nearest known blazar, has historically flared to more than 10 times the flux of the Crab nebula~\cite{bib:tluczykont}, and recently exhibited an episode of flaring more intense than any other previously observed flare in this source~\cite{bib:atel}.
We now outline some general considerations in the characterization of blazar flares and the possibility of detecting IGMF-induced delayed cascade emission following a flare.

\subsection*{Flare Model}

\begin{figure}[t!]
\centering
\includegraphics[width=0.45\textwidth]{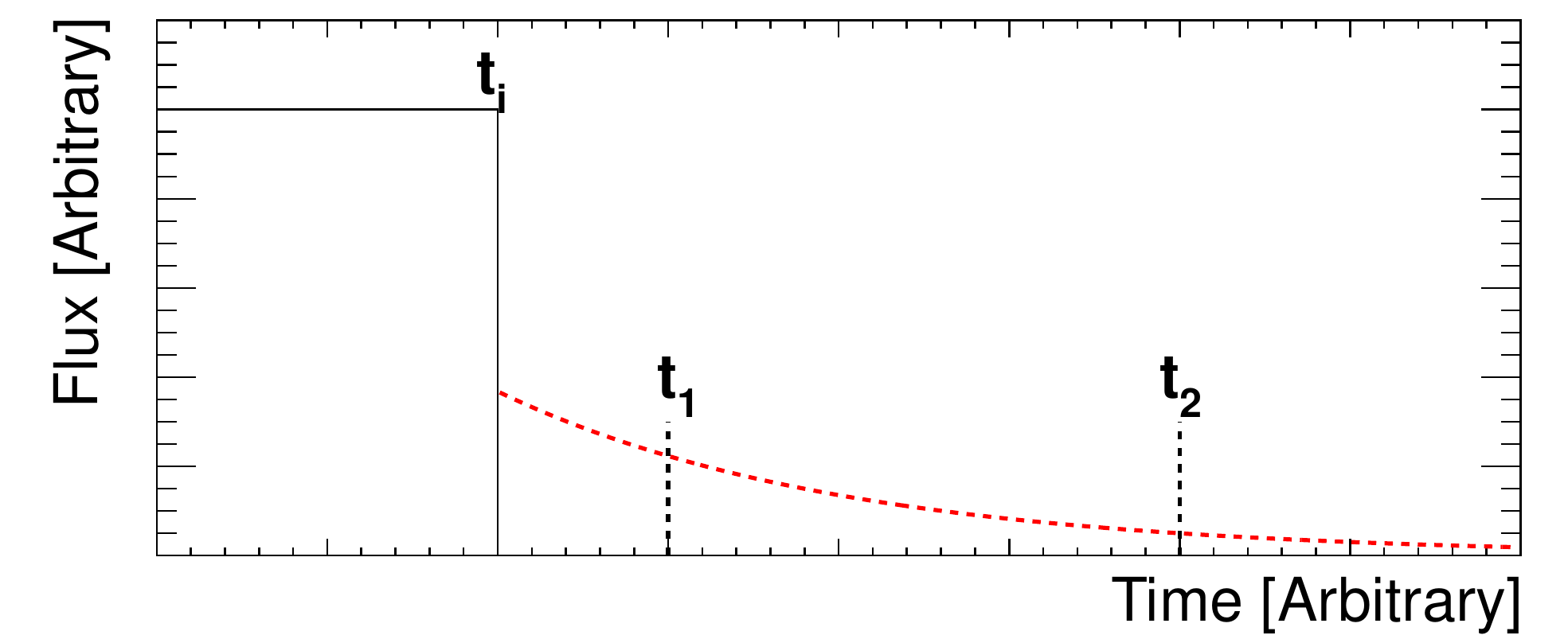}
\caption{A conceptual model of the light curve for a blazar flare of duration $t_i$ (see text). The intrinsic flare is shown as a solid black curve, while the delayed secondary appears as a red dashed curve. Observations of the secondary flux occur between times $t_1$ and $t_2$.}
\label{fig:flare}
\end{figure}

We assume a simple model in which the blazar emits a flare with a constant intrinsic flux $F_i$ that persists for a time $t_i$.
Figure~\ref{fig:flare} depicts a conceptual model of this flare followed by the secondary flux that arises due to interactions with the EBL and CMB.
If we observe this secondary flux between times $t_1$ and $t_2$, we can write the average value of this flux as
\begin{equation}
\label{eqn:secondary_flux}
F_s=\frac{f}{1-f}I(t_1,t_2)\frac{t_i}{t_2-t_1}F_i,
\end{equation}
where $f$ is the ratio of the number of secondary gamma rays to the total number of primary and secondary gamma rays, and
\begin{equation}
\label{eqn:norm}
I(t_1,t_2)=\frac{\int_{t_1}^{t_2}dt'g(t')}{\int_{0}^{\infty}dt'g(t')}
\end{equation}
is the normalized integral of the secondary flux $g(t)$ between times $t_1$ and $t_2$.

The spectrum of the intrinsic flare in our model takes the form
\begin{equation}
\label{eqn:intrinsic_spectrum}
F(E)=F_0E^{-\gamma}\exp(-E/E_C),
\end{equation}
where $F_0$ is a normalization parameter, $\gamma$ is the spectral index, and $E_C$ is an exponential cutoff energy.

\subsection*{Detectability Requirements}

For background-limited observations, the significance $S$ of the secondary detection, expressed in $\sigma$, will scale with the product of the secondary flux $F_s$ and the square root of the observation time:
\begin{equation}
\label{eqn:secondary_significance_start}
S=S_0F_s\sqrt{t_2-t_1},
\end{equation}
with $S_0$ an instrument-dependent scaling parameter expressed in flux units per $\sqrt{\textnormal{time}}$.
Plugging in equation~\ref{eqn:secondary_flux}, we find
\begin{equation}
\label{eqn:secondary_significance}
S=S_0\frac{f}{1-f}I(t_1,t_2)\frac{t_i}{\sqrt{t_2-t_1}}F_i.
\end{equation}
The detectability of the delayed emission, then, depends on the scale of the delay and the fraction of cascade gamma rays, both of which are energy-dependent quantities.

According to equation~\ref{eqn:secondary_significance}, the cascade fraction $f$ should be as large as possible to maximize the significance of the detection.
The optimum observation time $t_2-t_1$ is dictated by two factors.
First, $t_1$ must be long enough after the flare's end to avoid confusing the delayed flux with the intrinsic flux.
If the flare starts at $t=0$, then $t_1=t_i$ in the ideal case.
Second, $t_2$ should be selected such that $I(t_1,t_2)(t_2-t_1)^{-1/2}$ is maximized.
That is, as $I(t_1,t_2)$ approaches $1$, further observation becomes disadvantageous as the significance decreases according to the square root scaling of equation~\ref{eqn:secondary_significance}.

\section*{Simulated Predictions}

We use a dedicated simulation of the propagation of gamma rays, electrons, and positrons through the intergalactic medium to determine the detectability of flares with HAWC and other instruments.
The simulation includes interactions with the CMB, EBL, and IGMF.
Here, we use the simulation to investigate the cascade fraction and the arrival times of gamma rays in the cascade.

\subsection*{Cascade Fraction}

\begin{figure}[t!]
\centering
\includegraphics[width=0.45\textwidth]{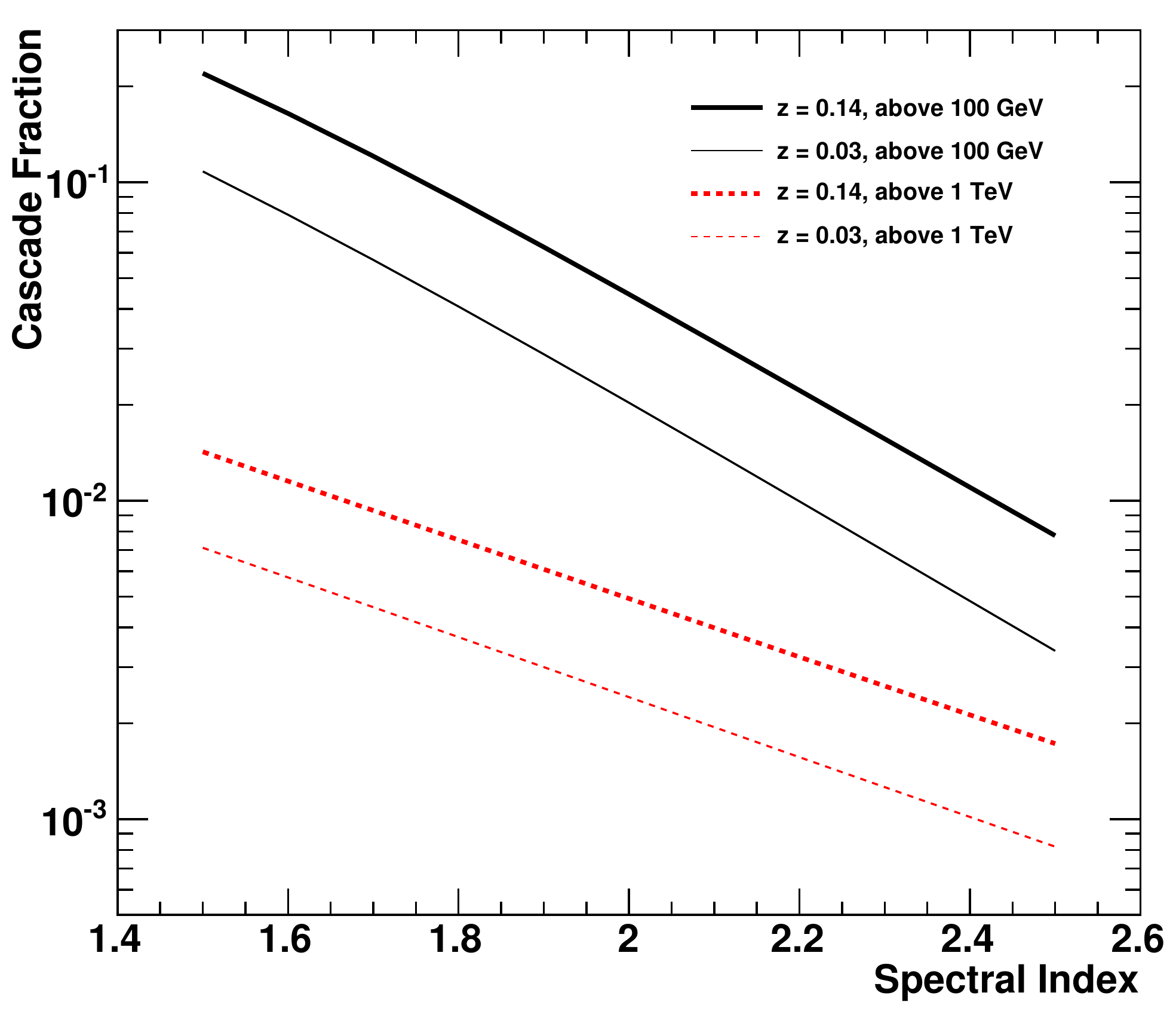}
\caption{Fraction of gamma rays in the cascade as a function of the spectral index of the intrinsic flare. The thick lines are for a redshift of 0.14, while the thin lines are for a redshift of 0.03. The solid black lines indicate the integral above 100 GeV, while the dashed red lines are for the integral above 1 TeV. We assume a cutoff energy of 5 TeV.}
\label{fig:index}
\end{figure}

Figure~\ref{fig:index} shows the value of the cascade fraction $f$ for two different redshifts, $z=0.03$, corresponding to Mrk421, and $z=0.14$, corresponding to a more distant blazar.
Curves for energies above 100 GeV and above 1 TeV appear in the figure.
Although HAWC is sensitive to energies as low as 50 GeV, the sensitivity changes rapidly below 1 TeV, and we expect the actual value of $f$ to lie somewhere between the two curves.

\begin{figure}[t!]
\centering
\includegraphics[width=0.45\textwidth]{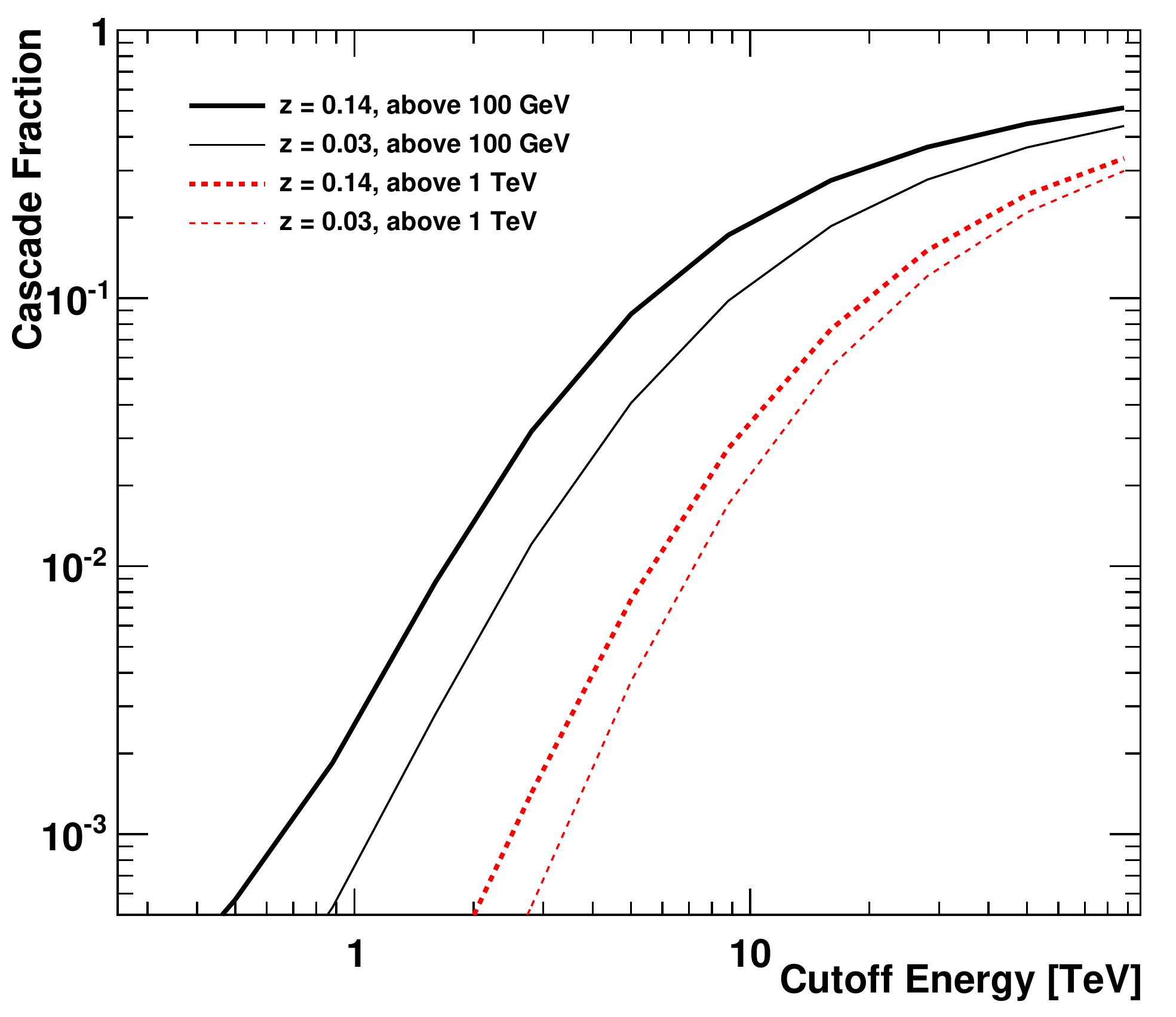}
\caption{Fraction of gamma rays in the cascade as a function of the cutoff energy of the intrinsic flare. The curves are the same as in figure~\ref{fig:index}.}
\label{fig:cutoff}
\end{figure}

In figure~\ref{fig:cutoff}, the same curves appear as a function of the cutoff energy $E_C$, for a fixed spectral index of $\gamma=1.8$.
It is clear from this figure that the fraction of gamma rays in the cascade is very sensitive to the presence of multi-TeV gamma rays in the intrinsic flare.
Combining the results from figures~\ref{fig:index} and~\ref{fig:cutoff}, we see that for a flare with a cutoff energy of 5 TeV and a particularly hard intrinsic spectrum, it is probably reasonable to assume that the number of gamma rays in the cascade represents a few percent of the total in the HAWC energy range.
Therefore, only a few extreme flares will have properties favorable for IGMF detection with HAWC.
However, because of its wide field of view, HAWC will be able to catch many more flares than IACTs, boosting the chances of locating a hard flare with a high cutoff energy.

In the Fermi energy band, the picture is much different.
Although equation~\ref{eqn:secondary_significance} no longer applies because the flare and secondary emission occur in different energy bands, the cascade flux will be dominant, yielding a value of $f$ very close to $1$.
Thus, the interpretation of HAWC and Fermi data collected simultaneously during a flare may well yield information about the IGMF that would not be accessible to either instrument alone.

\subsection*{Time Scale}

The time scale over which the cascade flux is observed should be long enough not to be confused with the intrinsic flare, but short enough that the significance as given by equation~\ref{eqn:secondary_significance} is not reduced too much.
We first consider the flare properties necessary for a detection of the IGMF using HAWC observations alone.
When fully constructed, HAWC will detect the Crab nebula at a significance of $5\sigma$ in a single transit.
It makes sense, then, to measure fluxes in Crab units and time in transits, or days.
In this case, we can select some reasonable values for the parameters of equation~\ref{eqn:secondary_significance}, $S_0\approx5$, $f\approx0.05$, and $I(t_1,t_2)\approx0.9$.
Plugging these values in, we arrive at
\begin{equation}
\label{eqn:hawc_estimate}
F_it_i\approx20\sqrt{t_2-t_1}.
\end{equation}
In other words, for a cascade with a time scale of 100 days, if the flare lasted for 10 days, it would need to be about 20 Crab units.
Although such an extreme event is rather unlikely, it is worth noting that these simple calculations demonstrate that the sensitivity of HAWC is good enough to make a marginal detection of the delayed flux following a flare.
Moreover, equation~\ref{eqn:hawc_estimate} should be taken as an order-of-magnitude estimate rather than a firm limit, and it is likely that a more detailed analysis could reveal additional sensitivity.

\begin{figure}[t!]
\centering
\includegraphics[width=0.45\textwidth]{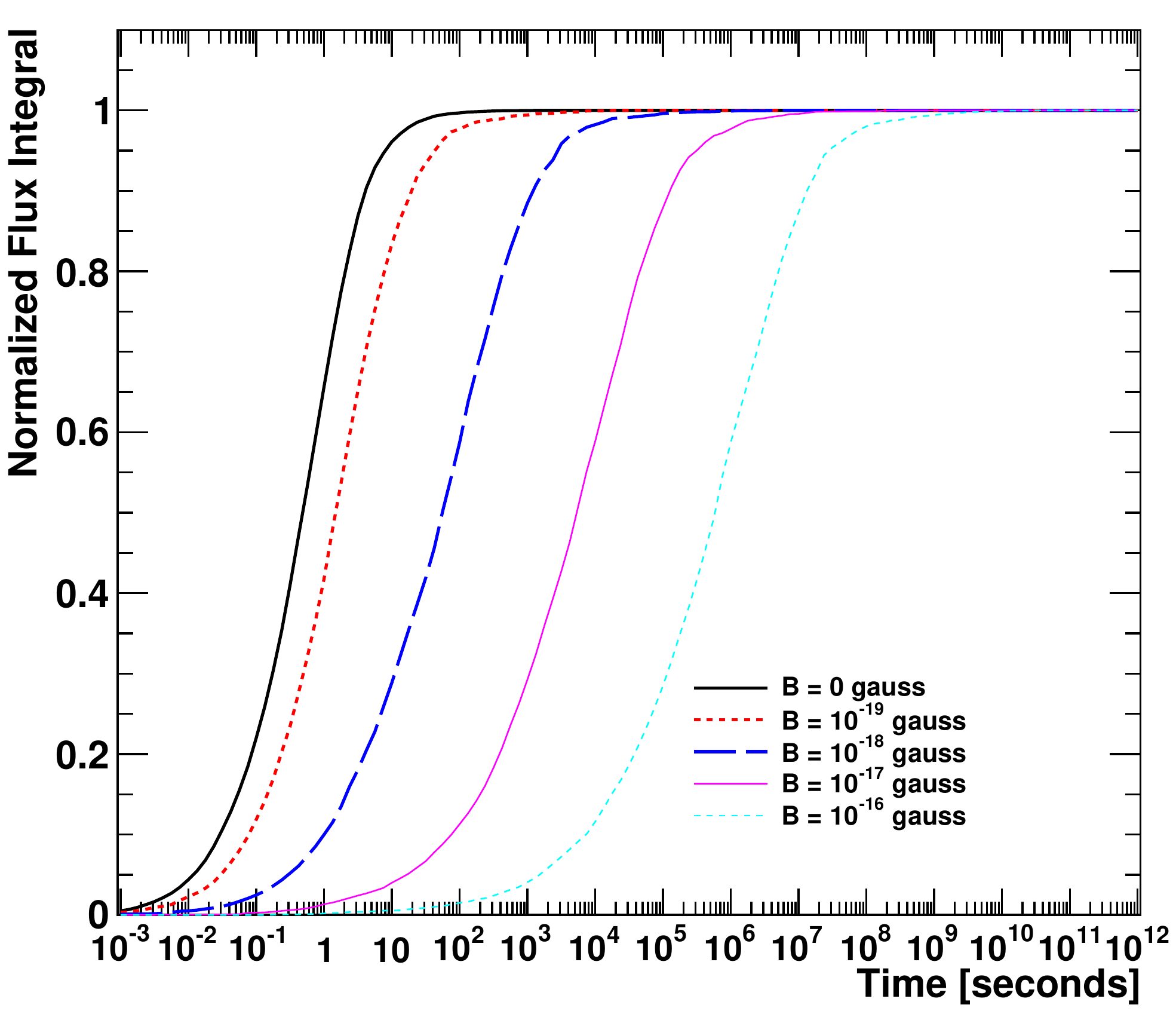}
\caption{The cumulative distribution of gamma rays in the cascade as a function of time. The intrinsic flare was injected at the redshift of Mrk421 with a spectral index of $\gamma=1.5$ and a cutoff energy of 10 TeV. The curves are for field strengths ranging from 0 gauss to $10^{-16}$ gauss.}
\label{fig:hawc_times}
\end{figure}

Figure~\ref{fig:hawc_times} displays the cumulative distribution of gamma rays in the cascade arising from an intrinsic flare of Mrk421.
By simple geometry arguments~\cite{bib:huan}, we know that the delay time should scale with the square of the field strength.
The curves indicate the time scale over which the cascade develops to some fraction of its total value, and readily demonstrate the expected scaling.
For an IGMF of zero strength, the opening angles of the inverse Compton and pair production processes cause a delay time of a few seconds, while for $B=10^{-16}$ gauss it takes about a year for the cascade to develop fully.
If the intrinsic flare lasts for a week, then a field strength between $10^{-17}$ and $10^{-16}$ gauss will induce cascade delays on the order of a few months, much longer than the time scale of the flare, so HAWC should be sensitive to an IGMF with a strength in this range.

We next turn to the combination of HAWC and Fermi observations.
For low field strengths, the cascade emission will be dominant in the Fermi energy band, and any time scale that is less than a year but longer than the intrinsic flare is likely to be detectable.
Furthermore, it may be possible via a sophisticated analysis of the flare structure to rule out a characteristic decay time scale in the Fermi data.
In this case, we would be able to place a firm lower limit on the strength of the IGMF.

\begin{figure}[t!]
\centering
\includegraphics[width=0.45\textwidth]{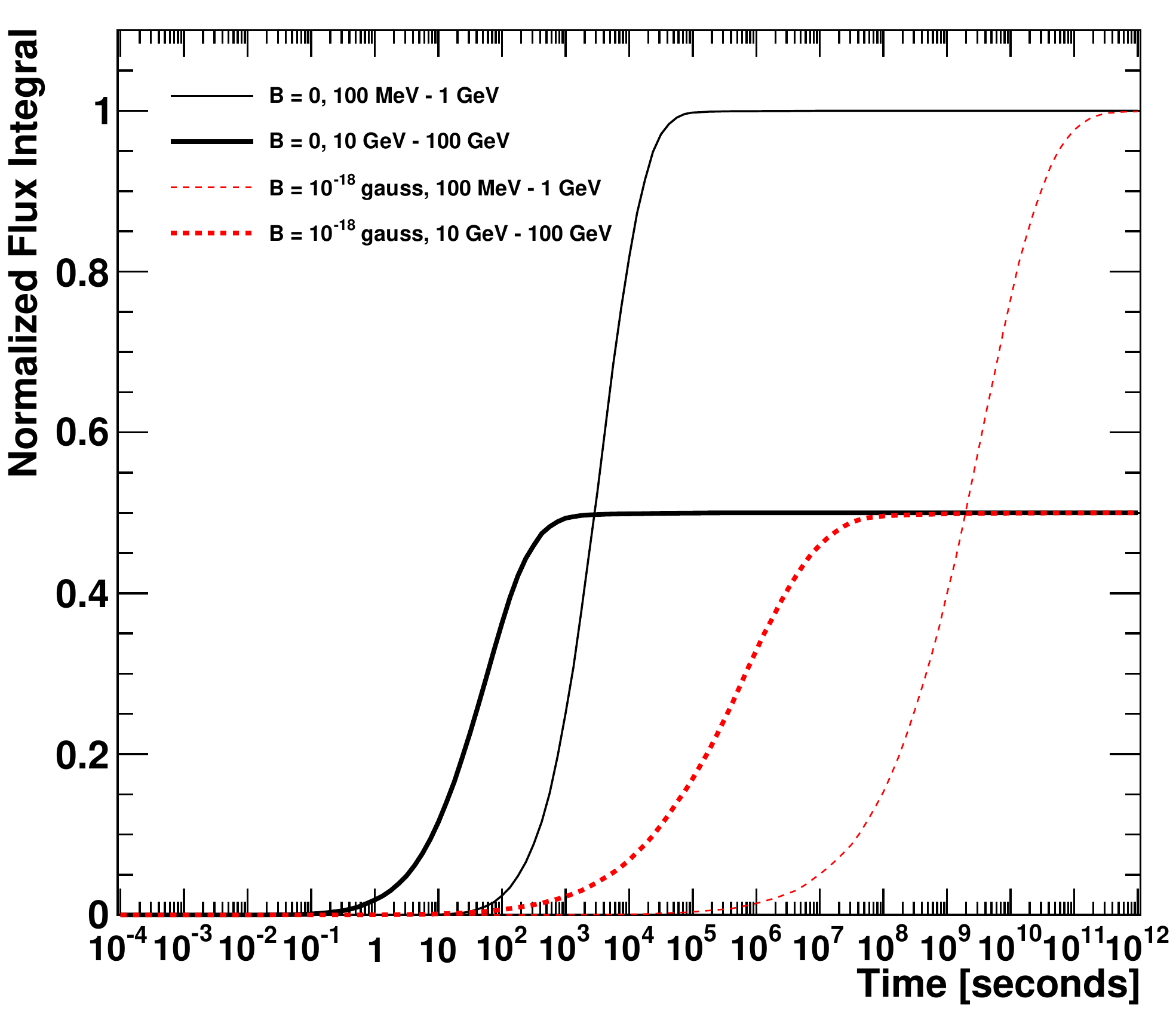}
\caption{Same as figure~\ref{fig:hawc_times}, but for two different energy bands in the Fermi sensitivity range and only two IGMF strengths. The thin curves are for gamma rays with energies between 100 MeV and 1 GeV, while the thick curves are for the 10 GeV to 100 GeV range. The solid black curves represent the case $B=0$; the dashed red curves represent $B=10^{-18}$ gauss. We scale the curves in the higher energy band by a factor of 0.5 for clarity.}
\label{fig:fermi_times}
\end{figure}

Figure~\ref{fig:fermi_times} shows the cumulative distribution of cascade gamma rays for field strengths of 0 and $10^{-18}$ gauss, and for two different energy ranges to which Fermi is sensitive.
Curves in the higher energy band are scaled by $0.5$ for clarity.
In the figure, it is clear that IGMF strengths around $10^{-18}$ gauss could be probed if the flare persists for no more than several days.
Additionally, in the lowest energy band, the time scale in the $B=0$ case is about an hour, suggesting that a careful study of the internal structure of an intrinsic flare would show decay features on hour time scales in the absence of an IGMF.
Conversely, the absence of such features would be strong evidence for the existence of a nonzero IGMF.

\section*{Detection Prospects}

In studying the IGMF by searching for cascade emission from blazars, it is crucial to have both TeV and GeV observations.
Once HAWC is fully operational, its ability to monitor all sources in its field of view without having to prioritize sources will render it very complementary to the IACTs and Fermi.
HAWC will contribute to the development of a firm and unbiased understanding of the average fluxes of gamma rays from blazars, informing studies of the IGMF.
Furthermore, HAWC will be able to detect blazar flares and identify those that are most promising for studies of the IGMF.

Using HAWC data alone, the detection of the IGMF in the gamma-ray signals from a flare may be possible if the strength is around $10^{-16}$ gauss.
By including data from Fermi, we can certainly extend the reach of this technique to $10^{-18}$ gauss.
With a detailed analysis of the time structure within the flare, we may be able to identify the absence of hour-long decays following flare sub-structure, which would provide strong evidence for the existence of the IGMF.

The IGMF may not be the only mechanism by which time delays are introduced into the gamma rays from a flare.
One other possible source would be Lorentz invariance violation (LIV), for which the propagation speed of the gamma rays would be energy dependent.
An LIV-induced delay, however, will apply to all gamma rays in the flare, whereas the IGMF signal appears only in the cascade.
Thus, a clear signature of the IGMF will be that only a fraction of the gamma rays in a particular energy band will be delayed relative to the direct flux from the flare.
Other mechanisms that introduce time delays can most likely be dismissed for the same reason.

In the future, HAWC will play a major role in any detection of the IGMF via blazar flares.
Due to its ability to monitor the entire overhead sky, HAWC will catch flares as they happen, producing a large data set and identifying those flares most promising for a follow-up study using Fermi data.
In the longer term, an instrument similar to HAWC but with significantly improved sensitivity could probe a broader range of IGMF strengths because it would be sensitive to delayed emission following weaker flares.
HAWC can function as a pathfinder for such an experiment.

\section*{Acknowledgements}

We acknowledge the support from: US National Science Foundation (NSF); US Department of Energy Office of High-Energy Physics; The Laboratory Directed Research and Development (LDRD) program of Los Alamos National Laboratory; Consejo Nacional de Ciencia y Tecnolog\'{\i}a (CONACyT), M\'exico; Red de F\'{\i}sica de Altas Energ\'{\i}as, M\'exico; DGAPA-UNAM, M\'exico; and the University of Wisconsin Alumni Research Foundation.

\clearpage

%\end{document}

\end{document}